\documentclass[12pt]{article}

\usepackage{amstext}
\usepackage{latexsym}
\usepackage{amssymb}
\usepackage{amsmath}
\def\R{\mathbb R}

\def\N{\mathbb N}
\allowdisplaybreaks

\newcommand{\intkl}{\int_{|z|\leq ct}}
\newcommand{\intgl}{\int_{|z|=ct}}
\newcommand{\intgr}{\int_{|z|\ge ct}}
\newcommand{\tdxzp}{(\hat{t}(z), x+z, p)}
\newcommand{\tdxz}{(\hat{t}(z), x+z)}
\newcommand{\dt}{\partial_t}
\renewcommand{\varphi}{\phi}

\begin{document}

\newtheorem{theorem}{Theorem}[section]
\renewcommand{\thetheorem}{\arabic{section}.\arabic{theorem}}
\newtheorem{definition}[theorem]{Definition}
\newtheorem{deflem}[theorem]{Definition and Lemma}
\newtheorem{lemma}[theorem]{Lemma}
\newtheorem{example}[theorem]{Example}
\newtheorem{remark}[theorem]{Remark}
\newtheorem{remarks}[theorem]{Remarks}
\newtheorem{cor}[theorem]{Corollary}
\newtheorem{pro}[theorem]{Proposition}
\newtheorem{proposition}[theorem]{Proposition}

\renewcommand{\theequation}{\thesection.\arabic{equation}}

\title{Post-Newtonian  Approximation \\
       of the Vlasov-Nordstr{\"o}m System}
\author{{\sc Sebastian Bauer\footnote{Supported in parts by
        DFG priority research program SPP 1095}} \\[2ex]
        Universit\"at Duisburg--Essen, FB  Mathematik, \\
        D\,-\,45117 Essen, Germany \\[1ex]
        \\ \hspace{-4.3em} 1.5 post-Newtonian approximation, Darwin system}
\date{}
\maketitle
\begin{abstract}\noindent
We study the Nordstr{\"o}m-Vlasov system which describes the dynamics of a
self-gravitating ensemble of collisionless particles in the framework of the
Nordstr{\"o}m scalar theory of gravitation. If the speed of light $c$ is
considered as a parameter, it is known that in the Newtonian limit $c\to\infty$
the Vlasov-Poisson system is obtained. In this paper we determine a higher 
approximation  and establish a pointwise error estimate of order ${\cal
  O}(c^{-4})$. Such an approximation is usually called a 1.5 post-Newtonian approximation.  
  
\end{abstract}


\setcounter{equation}{0}

\section {Introduction and Main Results}
In astrophysics stars of a galaxy are often modeled by a collisionless gas
interacting only by the gravitational fields which they  create
collectively; together this leads to the Vlasov-Einstein system. Recently, interest has
arisen in a simplified but still relativistic model essentially going back
to Nordstr{\"o}m, see \cite{Nord}, in which the dynamic of the matter is coupled to a
\textit{scalar} theory of gravitation, the metric tensor  potential is replaced by a
scalar function and Einstein's equations are replaced by a wave equation. In \cite{carein}
a reformulated version of Nordstr{\"o}m's theory is presented in which the
system reads
\begin{equation}\label{VN}\tag{VN{\tt{c}}}
  \left.
   \begin{array}{c}\displaystyle 
    S(f)-\Big[ S(\varphi)p+\gamma c^2\nabla_x\,\varphi\Big]\cdot\nabla_p\,f 
    = 4 S(\varphi) f                                                  
     \\ \displaystyle  
    \mu = \int \gamma f \,dp                                      
     \\ \displaystyle 
    -\partial_t^2\varphi + c^2 \Delta_x\, \varphi = 4 \pi \mu.
    \end{array}
   \right\}
\end{equation}
In the previous equations, $f=f(t, x, p)$ gives the probability density to find
a particle at time $t$ at position $x$ with momentum $p$ where $t\in\R,\;x\in\R^3,\;p\in\R^3$.
$\varphi=\varphi(t, x)$ is the mean Nordstr{\"om} gravitational potential
generated by the particles, $c$ the speed of light. Moreover
\begin{equation*}
   p^2 = |p|^2,\quad
   \gamma=(1 + c^2 p^2)^{-1/2}, \quad
   \hat p = \gamma p, \quad\text{and}\quad 
   S = \partial_t + \hat p \cdot \nabla_x
\end{equation*}
is the relativistic free-streaming operator and $\hat{p}$ the relativistic
velocity associated to the momentum $p$.

Together with the initial conditions
\begin{equation}
   f(0, x, p) = f^\circ (x, p), \quad \varphi (0, x) = \varphi^0(x), \quad \partial_t \varphi (0, x) = 
   \varphi^1 (x) \label{IC}
\end{equation}
(\ref{VN}) is the Cauchy problem of the Vlasov-Nordstr{\"o}m system.
For a physical interpretation and a derivation of this system see \cite{calee}. 

In this formulation (\ref{VN}) exhibits many similarities to the relativistic
Vlasov-Maxwell system, which also consists of a, in that case homogenous, Vlasov
equation coupled to a linear hyperbolic equation. Thus, many  techniques
developed for the Vlasov-Maxwell system also apply for the
Vlasov-Nordstr{\"o}m system providing existence and uniqueness of local classical
solutions  as well as a continuation criterion, which guarantees that the
solution is global in time if a control for the velocity support of the matter
density is established, see \cite{carein} . Furthermore, there are results about existence of global weak solutions, see
\cite{carein-II}, existence of global classical solutions in the 2D case, see
\cite{lee-II}, for spherically symmetric initial data,  see \cite{andcalrein},
and for small initial data, see \cite{fried}. 
Concerning the
Vlasov-Maxwell system we refer the reader to \cite{glassey:96} and the
references cited therein. On the other hand
the same important pieces of the basic theory are missing, namely global existence
of classical solutions with unrestricted initial data and uniqueness of
weak solutions. However, it is remarkable that there is a blow-up result for
(\ref{VN}) if the sign in the force-term of the Vlasov equation is changed,
see \cite{carein}.    

This paper is concerned with the non-relativistic limit of (\ref{VN}), i.e. the
limit $c\to\infty$: From physical intuition the Vlasov-Poisson system is the
non-relativistic limit of the Vlasov-Einstein system, a statement which has
been made rigorous under certain circumstances in \cite{ADR}.  
In \cite{calee} it has  been shown that as $c\to\infty$ also solutions of
(\ref{VN}) converge to a solution of a Vlasov-Poisson equation in a pointwise
sense obeying an error estimate of order ${\cal O}(c^{-1})$. Thus, (\ref{VN}) can be regarded as
another relativistic generalization of the Vlasov-Poisson system. 
The situation in the Maxwell case is similar as solutions of the
Vlasov-Maxwell system approach a solution of the Vlasov-Poisson system (in the
plasma case, which differs from the Vlasov-Poisson system for the
gravitational case by the sign in the force-term) in the same sense, see \cite {schaeffer:86}. 
    
It is the goal of this paper to replace  the Vlasov-Poisson system, the
classical or Newtonian limit,
by another effective equation to achieve higher order convergence and a more
precise approximation (in the plasma case this has been done in \cite{baukun}).  
This will lead to an effective system whose solution stay as close as ${\cal
  O}(c^{-4})$ to a solution of (\ref{VN}) if the initial data are matched
appropriately. Thus, it is the 1.5 Post-Newtonian approximation, note that in the
context of General Relativity the approximations are usually counted in powers
of $c^{-2}$.
 
In order to derive effective equations in the limit of small initial velocities 
we shall formally expand all quantities in powers of $ c^{-1} $
\begin{eqnarray*}
   f   &=& f_0 + c^{-1}f_1 + c^{-2} f_2 + \cdots \\
   \mu &=& \mu_0 + c^{-1} \mu_1 + c^{-2} \mu_2 + \cdots \\
   \varphi &=& \varphi_0 + c^{-1} \varphi_1 + c^{-2} \varphi_2 + \cdots 
\end{eqnarray*} 
put this ansatz into (\ref{VN}) and derive an equation in every order of $ c^{-1}$.

Therefore, we obtain in the orders $ c^2 $ and $ c^1 $
\begin{equation*}
   -\Delta_x\, \varphi_0 = 0, \quad -\Delta_x\, \varphi_1 = 0, 
\end{equation*}
thus, we set 
\begin{equation}\label{P0P1}
   \varphi_0 = \varphi_1 = 0 .
\end{equation}
In the zeroth order we obtain a Vlasov-Poisson System
\begin{equation}\label{VP}
   \left.\begin{array}{c}\displaystyle 
   \tilde S(f_0) - \nabla_x\, \varphi_2 \cdot \nabla_p f_0 = 0 ,\qquad
   \mu_0 = \int f_0\,dp,                                     
   \\
   \varphi_2 (t, x) = \displaystyle -\int |z|^{-1} \mu_0 (t, x+z)\, dz,  \qquad
   f_0(0, x, p)  =  f^\circ (x, p)
   \end{array}\right\}\tag{VP}                               
\end{equation}
where
\begin{equation*}
   \tilde S = \partial_t + p \cdot \nabla_x
\end{equation*}
is the non-relativistic free-streaming operator.
In the first order a linearized Vlasov-Poisson system appears
\begin{equation*}
   \left.\begin{array}{c}
   \tilde S (f_1) - \nabla_x\, \varphi_3 \cdot \nabla_p\, f_0 
   -\nabla_x\, \varphi_2 \cdot \nabla_p\, f_1 = 0,
   \\ 
   \mu_1 = \displaystyle \int f_1 \,dp,\qquad
   \Delta_x\, \varphi_3 =  \mu_1 + \partial_t^2 \varphi_1. 
   \end{array}\right\}
\end{equation*}  
Hence, if we suppose that $ f_1 (0,  x, p) = 0 $,
we can set 
\begin{equation*}
   f_1 = 0 \quad \text{and} \quad \varphi_3 = 0,
\end{equation*}
which also yields
\begin{equation}
   \mu_1 = 0.
\end{equation}
In the second order we derive an inhomogeneous Vlasov equation 
coupled to a Poisson equation
\begin{equation}\label{LVP}
   \left.\begin{array}{c}\displaystyle
   \tilde S(f_2)-\nabla_x\,\varphi_2\cdot\nabla_p\,f_2 
   -\nabla_x\,\varphi_4\cdot\nabla_p\,f_0=  
   \\ \displaystyle
    4f_0\tilde S(\varphi_2)+\frac{p^2}{2}p\cdot \nabla_x f_0
     +\Big(\tilde S(\varphi_2)p-\frac{p^2}{2}\nabla_x\varphi_2\Big)\cdot\nabla_p\,f_0
   \\ \displaystyle
   \mu_2  =  \int(f_2 -1/2p^2 f_0 )\,dp,\qquad     
   \Delta_x\,\varphi_4  =  \mu_2+\partial_t^2\varphi_2     
   \end{array}\right\}\tag{LVP}
\end{equation}
for which we choose homogeneous initial data
\begin{equation}\label{IC-LVP}
   f_2(0, x, p) = 0.
\end{equation}
At this point we need to discuss the Poisson equation for $\varphi_4$. Since 
$\Delta_x \partial_t^2 \varphi_2 = 4\pi\partial_t^2\mu_0$ the Poisson equation
in (\ref{LVP}) can
be rewritten as 
\begin{equation}\label{Poisson2}
   \varphi_4 = \Delta_x^{-1}(4\pi\mu_2) +
   (\Delta_x )^{-2}(4\pi\partial_t^2\mu_0).
\end{equation}
Therefore, we define $\varphi_4$ by
\begin{equation}\label{phi4-def}
   \varphi_4(t, x) = -\int\int|z|^{-1}\big(f_2-1/2p^2f_0\big)(t, x+z,
   p)\,dp\,dz-\frac{1}{2}\int|z|\partial_t^2f_0(t, x+z, p)\,dp\,dz.
\end{equation}
Using (\ref{VP}) and partial integration the second term can be rewritten as 
\[ -\frac{1}{2}\int|z|\partial_t^2 f_0(t, x+z, p)\,dp\,dz =
-\frac{1}{2}\int\int(\bar{z}\cdot p)\dt f_0(t, x+z, p)\,dp\,dz,\qquad \bar{z}=z|z|^{-1}.
\]
Thus far this term is merely  bounded instead of decaying  at infinity 
and  $\varphi_4$ seems to be  determined only
modulo a function  depending on $t$. But if the boundary condition is fixed in
terms of integrability in weighted Sobolev spaces (for a definition see
section \ref{eff-syst}, below
(\ref{TrLVP})) instead of a pointwise
estimates, the corresponding condition for vanishing at infinity is 
\begin{equation}
  \label{int}
  \varphi_4\in L^p_s\qquad\text{for every $s<-2+3/p'$}
\end{equation} 
where  $1<p<\infty,\;1/p+1/p'=1$.  We will
show in section \ref{eff-syst} that in fact $\varphi_4$ fulfills this
integrability condition.

The first aim of this paper is to show that 
\begin{eqnarray}\label{Darwin-def}
   f^D  & := &  f_0+c^{-2}f_2, 
   \\ \nonumber
   \varphi^D & = & c^{-2}\varphi_2+c^{-4}\varphi_4
\end{eqnarray}
yields a higher order pointwise approximation of the Vlasov-Nordstr{\"o}m than
the  Vlasov-Poisson  system. We will call this system the Darwin system  because in the
case of individual charged particles the relevant approximation is usually called the
Darwin approximation. 
It is clear that for achieving this improved approximation property also the
initial data of the Vlasov-Nordstr{\"o}m model has to be matched appropriately
by the data for the  Darwin system. For a prescribed initial density $f^\circ$
we are able to calculate $(f_0, \varphi_2)$ and $(f_2, \varphi_4)$ according
to what has been outlined above. We then consider the Vlasov-Nordstr{\"o}m
system with the initial condition
\begin{eqnarray}\label{IC-matched}
    f(0, x, p) & = & f^\circ(x, p)
    \\ \nonumber
    \varphi^0(x) =\varphi(0, x) & = &  c^{-2}\varphi_2(0, x)+c^{-4}\varphi_4(0, x) 
    \\ \nonumber
    \varphi^1(x) =\partial_t\varphi(0, x) & = &  c^{-2}\partial_t\varphi_2(0,
    x)+c^{-4}\dt\varphi_4(0, x).
\end{eqnarray}
Before we formulate the main theorem of this paper let us recall that
solutions of the Vlasov-Nordstr{\"o}m system together with the initial
conditions (\ref{IC-matched}) exist at least on some time interval $[0, T]$
which is independent of $c\geq 1$; see \cite[Thm. 3]{calee}. This time
interval is fixed throughout this paper.
\begin{theorem}\label{Hauptsatz}
Assume that $f^\circ\in C^\infty(\R^3\times\R^3)$ is nonnegative
and has compact support. From $f^\circ$ calculate $(f_0, \varphi_2)$ and
$(f_2, \varphi_4)$, and then define initial data for the
Vlasov-Nordstr{\"o}m system
by (\ref{IC-matched}). Let $(f, \varphi)$ denote the solution of the
Vlasov-Nordstr{\"o}m system 
with initial data (\ref{IC-matched}) and let $(f^D, \varphi^D)$ be defined
as in (\ref{Darwin-def}). Then there exists a constant $M>0$,
and also for every $R>0$ there is $M_R>0$, such that
\begin{eqnarray}
   |f(t, x, p)-f^D(t, x, p)| & \leq & Mc^{-4}\quad\hspace{0.55em} (x\in\R^3), 
   \nonumber \\
   |\varphi(t, x)-\varphi^D(t, x)| & \leq & M_R\,c^{-4}\quad (|x|\le R), \label{diff-esti} \\
   |\partial_t\varphi(t, x)-\partial_t\varphi^D(t, x)| & \leq &
   Mc^{-4}\quad\hspace{0.65em} (x\in\R^3), 
   \nonumber\\
   |\nabla_x\varphi(t, x)-\nabla_x\varphi^D(t, x)| & \leq &
   M_R\,c^{-6}\quad(|x|\le R)\nonumber
\end{eqnarray}
for all $p\in\R^3$, $t\in [0, T]$ and $c\geq 1$.
\end{theorem}
The constants $M$ and $M_R$ are independent of $c\geq 1$ but do depend on the initial data. Note that if the
Vlasov-Nordstr{\"o}m system is compared to the Vlasov-Poisson system only, one
obtains the estimates $|f(t, x, p)-f_0(t, x, p)|+c^2|\varphi(t,
x)-\varphi_2(t, x)|+c^2|\nabla_x\varphi(t,
x)-\nabla_x\varphi_2(t, x)|+|\dt\varphi(t, x)|\leq M c^{-1}$; see \cite[Thm. 3]{calee}.

While this approximation has the big advantage that, since by now the
Vlasov-Poisson system is well understood, the existence of $(f_0, \varphi_2)$
and also of $(f_2, \varphi_4)$ does not pose serious problems; note that in
(\ref{LVP}) the equation for $f_2$ is linear. Therefore one can hope to get
more information on (\ref{VN}) by studying the approximate equations.

As a drawback of the above hierarchy one has to deal with two densities
$f_0,\;f_2$  and two fields $\varphi_2$ and $\varphi_4$ to define $f^D$. Thus
it is natural to look for a model which can be written down using only one density and
one field. The most natural candidate for such a system might be the
following.
  \begin{equation}
    \left.
     \begin{array}{c}\displaystyle
          \dt f+p\Big(1-\frac{p^2}{2c^2}\Big)\cdot \nabla_x f-
          \Big[\tilde{S}(\varphi)p+c^2\Big(1-\frac{p^2}{2c^2}\Big)\nabla_x\varphi\Big]\cdot \nabla_p f 
          =4\tilde{S}(\varphi)f  
         \\ \displaystyle
         \mu = \int\Big(1-\frac{p^2}{2c^2}\Big)f(\cdot, \cdot, p)\,dp 
         \\ \displaystyle
         \varphi = \frac{4\pi}{c^2} \Delta^{-1}\mu+\frac{4\pi}{c^4}\Delta^{-2}\dt^2\mu
      \end{array}
     \right\}\nonumber
  \end{equation}
Note, that $(f^D, \varphi^D)$ solves this system up to an error of order
$c^{-4}$. But since second derivatives occur it is not clear which initial
conditions are to be posed. It turns out that it is more convenient to rewrite
this system in terms of derivatives of $\varphi$ then in terms of  
$\varphi$. To this end we introduce the scalar force field $\psi^\ast$
corresponding to $\dt\varphi$ and the vector field $E^\ast$ corresponding to
$\nabla\varphi$. This leads to the following system which we call Darwin
Vlasov-Nordstr{\"o}m system. 
 \begin{equation}
    \label{DVNc}\tag{DVN{\tt{c}}}
    \left.
      \begin{array}{c}\displaystyle
         \dt f^\ast +p\Big(1-\frac{p^2}{c^2}\Big)\cdot \nabla_x
         f^\ast-\Big[(\psi^\ast+p\cdot E^\ast)p+c^2\Big(1-\frac{p^2}{2c^2}\Big)E^\ast\Big]\cdot \nabla_p f^\ast
         \\ \displaystyle
          =4(\psi^\ast+p\cdot E^\ast)f^\ast
         \\ \displaystyle
         \mu^\ast =\int\Big(1-\frac{p^2}{2c^2}\Big)f^\ast(\cdot, \cdot, p)\,dp,
         \qquad j^\ast=\int pf^\ast(\cdot, \cdot, p)\,dp 
         \\ \displaystyle
          \Delta\psi^\ast = -\frac{4\pi}{c^2}\nabla\cdot j^\ast,\qquad c^2\Delta E^\ast =
          4\pi\nabla\mu^\ast+\dt\nabla\psi^\ast
       \end{array}
    \right\} 
 \end{equation} 
Again $(f^D, \dt\varphi^D, \nabla_x\varphi^D)$ solves (\ref{DVNc}) up to an
error of order $c^{-4}$.
\begin{theorem}
   \label{DVN-thm} 
    Assume that $f^\circ\in C^\infty(\R^3\times\R^3)$
    is nonnegative and has compact support. Then there exist $c^\ast\ge 1$ and $T^\ast>0$
    such that the following holds for $c\geq c^\ast$.
    \begin{itemize}
        \item[(a)] If there is a local solution of (\ref{DVNc}), then the initial data
                   $\psi^\circ$ and $E^\circ$  of (\ref{DVNc}) at $t=0$ are uniquely determined
                   by the initial density $f^\circ$.
        \item[(b)] The system (\ref{DVNc}) has a unique $C^2$-solution
                   $(f^\ast, \psi^\ast, E^\ast)$ on $[0, T^\ast]$ attaining that initial data
                   $(f^\circ, \psi^\circ, E^\circ)$ at $t=0$. 
        \item[(c)] Define $\varphi^\ast$ by
                   \begin{equation}\label{phiast-def}
                      \varphi^\ast =
                      \frac{4\pi}{c^2}\Delta^{-1}\mu^\ast-\frac{4\pi}{c^4}\Delta^{-2}\big(\nabla\cdot
                      \dt j^\ast\big).
                   \end{equation}
                   Then $c^2\Delta\varphi^\ast = 4\pi\mu^\ast+\dt\psi^\ast$
                   and $\nabla_x\varphi^\ast= E^\ast$ as well as
                   $\dt\varphi^\ast=\psi^\ast+{\cal O}(c^{-4})$.
                                          
                 Let $(f, \varphi)$ denote the solution of (\ref{VN})
                 with initial data $(f^\circ, \varphi^\ast(0, \cdot),
                 \dt\varphi^\ast(0, \cdot))$. Then there exists
                 a constant $M>0$, and also for every $R>0$ there is $M_R>0$,
                 such that 
                \begin{eqnarray*}
                |f(t, x, p)-f^\ast(t, x, p)| & \leq & Mc^{-4}\quad\hspace{0.55em} (x\in\R^3), \\
                |\nabla_x\varphi(t, x)-E^\ast(t, x)| & \leq & M_R\,c^{-6}\quad (|x|\le R), \\
                |\dt\varphi(t, x)-\psi^\ast(x, t)| & \leq & Mc^{-4}\quad\hspace{0.6em} (x\in\R^3),\\
                |\varphi(t, x)-\varphi^\ast(t, x)| & \leq & M\,c^{-4}\quad\hspace{0.6em} (x\in\R^3),
                \end{eqnarray*}
                for all $p\in\R^3$, $t\in [0, \min\{T, T^\ast\}]$, and $c\geq
                c^\ast$. 
    \end{itemize}
\end{theorem}
We shall  continue with a few remarks about the Darwin approximation of
the relativistic Vlasov-Maxwell system derived in \cite{baukun}. In that paper only  the 1
PN approximation is treated but we want to emphasize that in the
case of only one species it is possible to establish the 1.5 PN approximation
for Vlasov-Maxwell as well in the same way as it is done for (\ref{VN}) in  this
paper.  On the other hand, if several species with different charge to mass
ratio, say $f^+$ with charge 1  and $f^-$ with charge -1, are under consideration, mass is set to unity
for both species, an effect due to radiation damping takes place, which is
proportional to 
  \begin{equation}
    \label{damping-i}
    c^{-3}\dt\int\int\nabla U_0(f_0^++f^-_0)(t, x, p)\,dp\,dx
  \end{equation}
where $(f^\pm_0, \,U_0)$ is a solution of the Vlasov-Poisson system in the
plasma case with two different species
\begin{equation}
  \label{VP-plasma}
  \left.
    \begin{array}{c}\displaystyle 
      \tilde{S}(f^\pm)\pm\nabla_x U_0\cdot\nabla_p f^\pm = 0
      \\ \displaystyle
      \Delta U_0 = 4\pi\int(f_0^+-f^-_0)\,dp.
    \end{array}
  \right\}\tag{VP{\tt{plasma}}}
\end{equation}
(In the case of only one species and in the gravitational case the
corresponding term vanishes, see (\ref{Rad1}) below.)
Note that $\dt\int\int\nabla U_0(f_0^++f^-_0)(t, x,
p)\,dp\,dx=\frac{d^3}{dt^3}D(t)$ where 
\begin{equation}
  \label{dipol-moment}
  D(t):=\int\int x(f^+_0-f^-_0)(t, x, p)\,dp\,dx
\end{equation}
is the dipole moment. 
Thus it seems reasonable to replace (\ref{VP-plasma}) by
\begin{equation}
  \label{VP-damped}
    \left.
    \begin{array}{c}\displaystyle 
      \tilde{S}(f^\pm)\pm\Big(\frac{a}{c^3}\frac{d^3}{dt^3}D(t)+\nabla_x U_0\Big)\cdot\nabla_p f^\pm = 0
      \\ \displaystyle
      \Delta U_0 = 4\pi\int(f_0^+-f^-_0)\,dp.
      \\ \displaystyle D(t)=\int\int x(f^+_0-f^-_0)(t, x, p)\,dp\,dx
    \end{array}
  \right\}\tag{VP{\tt{rad}}} 
\end{equation}
where $a\not=0$ is a constant which has to be calculated from the full system. 
However, in this system phase space has increased by 3 dimensions and it is
not so clear how to single out the initial data of $\frac{d^2}{dt^2}D$, note
that $D(0)$ and $\frac{d}{dt}D(0)$ are already determined by the initial data
of $f^\pm$. On a formal level (\ref{VP-damped}) can be reduced in several
ways, which are discussed in \cite{KR1} and \cite{KR2}.

In the gravitational case effects due to damping are expected to take place in
the 2.5 PN approximation and to  be connected with the fifth time derivative
of the quadrupole moment, see \cite{KR2}.  Thus, it seems reasonable that
approximations up to the order of 2PN 
conserve an energy which can be calculated by an expansion of the energy
\begin{equation}
  \label{energy-full}
  {\cal E}=c^2\int\int\sqrt{1+c^{-2}p^2}f(t, x,
  p)\,dp\,dx+c^2\int\Big(|\dt\varphi(t, x)|^2+c^2|\nabla_x\varphi(t, x)|^2\Big)\,dx,
\end{equation}
which is conserved by solutions of the full Nordstr{\"o}m system; 
note that  in the Vlasov-Maxwell case the corresponding statement is
true for approximations up to the 1PN level. However, here  this fails already in
the  Newtonian approximation in a rather  obscure manner; the
conserved energy of (\ref{VP}) is
$\int\int\frac{p^2}{2}f\,dp\,dx-\int|\nabla_x\varphi_2|^2\,dx$ whereas from
the expansion of (\ref{energy-full}) one would expect
$\int\int\frac{p^2}{2}f\,dp\,dx+\int|\nabla_x\varphi_2|^2\,dx$
to be conserved. But this is the energy of Vlasov-Poisson system in the plasma case.

The paper is organized as follows. Some facts concerning (\ref{VP}),
(\ref{LVP}), and (\ref{VN}) are collected in Section \ref{eff-syst} as well as
the proof of (\ref{int}).
The proof of Theorem \ref{Hauptsatz} is elaborated in Section \ref{HS-bew}.
Section \ref{DVN-bew} contains the proof of Theorem \ref{DVN-thm}.
For these proofs we will mostly rely on suitable representation formulas
for the fields (refined versions of those used in \cite{carein,calee}),
which are derived in the appendix, Section \ref{append}.
\smallskip

\noindent
{\bf Notation:} $B(0, R)$ denotes the closed ball in $\R^3$
with center at $x=0$ or $p=0$ and radius $R>0$.
The usual $L^\infty$-norm of a function $\varphi=\varphi(x)$ over $x\in\R^3$
is written as ${\|\varphi\|}_x$, and if $\varphi=\varphi(x, p)$,
we modify this to ${\|\varphi\|}_{x, p}$.
For $m\in\N$ the $W^{m, \infty}$-norms are denoted by ${\|\varphi\|}_{m, x}$, etc.
If $T>0$ is fixed, then we write
\[ g(t, x, p,  c)={\cal O}_{cpt}(c^{-m}), \]
if for all $R>0$ there is a constant $M=M_R>0$ such that
\begin{equation}\label{GOForm}
   |g(t, x, p, c)|\leq Mc^{-m}
\end{equation}
for $|x|\le R$, $p\in\R^3$, $t\in [0, T]$, and $c\geq 1$.
Similarly, we write
\[ g(t, x, p, c)={\cal O}(c^{-m}) \]
if there is a constant $M>0$ such that (\ref{GOForm}) holds for all
$x, p\in\R^3$, $t\in [0, T]$ and $c\geq 1$. In general, generic constants
are denoted by $M$.

\setcounter{equation}{0}
\section{Some properties of (\ref{VP}), (\ref{LVP}), and (\ref{VN})}
\label{eff-syst}

There is a vast literature on (\ref{VP}), see e.g.~\cite[Sect.~4]{glassey:96}
or \cite{rein} and the references therein. For our purposes we collect
a few well known facts about classical solutions of (\ref{VP}).

\begin{proposition} Assume that $f^\circ\in C^\infty(\R^3\times\R^3)$
is nonnegative and has compact support. Then there exists a unique global
$C^1$-solution $f_0$ of (\ref{VP}), and there are nondecreasing
continuous functions $P_{V\!P}, K_{V\!P}: [0, \infty[\to\R$ such that
\begin{gather}
   {\|f_0(t)\|}_{x, p}\leq {\|f^\circ\|}_{x, p}, \nonumber \\
   {\rm supp}\,f_0(t, \cdot, \cdot)
   \subset B(0, P_{V\!P}(t))\times B(0, P_{V\!P}(t)),
   \label{SchrankeGeschwVP} \\
   {\|f_0(t)\|}_{1, x, p}+{\|\varphi_2(t)\|}_{2, x}\leq K_{V\!P}(t), \nonumber
\end{gather}
for $t\in [0, \infty[$.
\end{proposition}

This result was first established by Pfaffelmoser
\cite{pfaffelmoser:92}, and simplified versions of the proof
were obtained by Schaeffer \cite{schaeffer:91} and Horst \cite{horst:93};
a proof along different lines is due to Lions and Perthame \cite{lions/perthame:91}.

For our approximation scheme we also need bounds on higher derivatives
of the solution. This point was elaborated in \cite{lindner:91}
where it was shown that if $f^\circ\in C^k(\R^3\times\R^3)$,
then $f_0$ possesses continuous partial derivatives w.r.t.~$x$ and $p$
up to order $k$. The existence of continuous time-derivatives then
follows from the Vlasov equation. Thus, $f_0$ and $\varphi_2$ are $C^\infty$
if $f^\circ$ is $C^\infty$, and by a redefinition of $K_{V\!P}$ we can assume that
\begin{equation}\label{SchrankeFeldVP}
   {\|f_0(t)\|}_{5, x, p}\leq K_{V\!P}(t),\quad t\in [0, \infty[.
\end{equation}

The existence of a unique $C^1$-solution $f_2$ of (\ref{LVP})
follows by a contraction argument, but we omit the details. Furthermore
it can be shown that there are nondecreasing continuous functions
$P_{LV\!P}, K_{LV\!P}: [0, \infty[\to\R$ such that
\begin{gather}
   \label{TrLVP}
   {\rm supp}\,f_2(\cdot, \cdot, t)
   \subset B(0, P_{LV\!P}(t))\times B(0, P_{LV\!P}(t)), \\
   \label{SchrankeLVP}
   {\|f_2(t)\|}_{2, x, p}+{\|\varphi_4\|}_{2, x}\leq K_{LV\!P}(t),
\end{gather}
for $t\in [0, \infty[$.

Next we shall investigate the boundary behavior of $\varphi_4$ in terms of
integrability in weighted Sobolev spaces where the weight function is defined
by $\rho(x)= (1+|x|^2)^{1/2}$. For $k\in \N_0,\;1\leq p<\infty$ and $s\in
\R$ we say $u\in W^{k, p}_s$ if and only if $u\in W^{k, p}_{\mathrm{loc}}(\R^3)$ and 
$\rho^{s+|\alpha|}\partial^\alpha u\in L^p(\R^3)$ for $0\leq |\alpha|\leq
k$; $\alpha\in\N_0^3$ a usual multi-index. In order to establish (\ref{int})
we use some mapping properties of the Laplacian in weighted Sobolev spaces due
to McOwen. Citing \cite{McO}, for $1<p<\infty,\;1/p+1/p'=1$,  we have that $\varphi_4\in
W^{2,p}_s$ for $s<-2+3/p'$ if $\dt^2\varphi_2\in L^p_{2+s}$. Furthermore,
$\dt^2\varphi_2\in L^p_{2+s}$ holds true if
\begin{equation}
  \label{ort}
  \Delta\dt^2\varphi_2=\dt^2\mu_0\in L^p_{4+s}\qquad\text{and}\qquad\int h(x)\dt^2\mu_0(x)\,dx = 0
\end{equation}
for every polynomial $h$ of degree less than 1.  Thus,  to prove (\ref{int})
it is sufficient to show (\ref{ort}).
Since $\dt^2\mu_0(t, \cdot)$ has compact support for every $t\geq 0$, compare
(\ref{SchrankeGeschwVP}), the first condition is clear. Now let
$h(x)=a_0+a_1x_1+a_2x_2+a_3x_3$, $ a_i\in\R$. Employing (\ref{VP}), some partial integrations
yield
  \begin{eqnarray*}
    \lefteqn{\int h(x)\dt^2\mu_0(t, x)\,dx = -\int\int
               h(x)p\cdot\nabla_x\dt f_0(t, x, p)\,dp\,dx}
    \\ & = & 
    \nabla h\cdot\int\int p\dt f_0(t, x, p)\,dp\,dx=\nabla h\cdot\int\int
    p(-p\cdot \nabla_xf_0+\nabla_x\varphi_2\cdot \nabla_p f_0)(t, x,
    p)\,dp\,dx
    \\ & = & 
    -\nabla h \cdot \int\nabla\varphi_2\mu_0(t, x)\,dx = -\frac{1}{4\pi}\sum_{i,j=1}^3
               a_i\int\partial_i\varphi_2\partial_j\partial_j\varphi_2(t, x)\,dx
   \\ & = & \frac{1}{8\pi}\nabla h\cdot \int\nabla|\nabla \varphi_2(x)|^2\,dx
               = 0.  
  \end{eqnarray*}

Concerning solutions of (\ref{VN}) we have 
the following from \cite[Thm.~3, Proposition 2]{calee}.
\begin{proposition}\label{Schaeffer-prop}
Assume that $f^\circ\in C^\infty(\R^3\times\R^3)$ is nonnegative
and has compact support. If $\varphi^0$ and $\varphi^1$ are defined by (\ref{IC-matched}),
then there exits $T>0$ (independent of $c$) such that for all $c\geq 1$
the system (\ref{VN}) with initial data (\ref{IC-matched}) has a unique $C^1$-solution
$(f, \varphi)$ on the time interval $[0, T]$. In addition, there are
nondecreasing continuous functions (independent of $c$)
$P_{V\!N}, K_{V\!N}: [0, T]\to\R$ such that
\begin{gather}
   f(x, p, t) = 0\quad\mbox{if}\quad |p|\geq P_{V\!N}(t),
   \label{SchrankeSupport} \\
   |f(t)|_{x ,p}+|\varphi(t)|_{1, x}\leq K_{V\!N}(t),
   \label{SchrankeFelder}
\end{gather}
for all $x\in\R^3$, $t\in [0, T]$, and $c\geq 1$.
\end{proposition}


\setcounter{equation}{0}

\section{Proof of Theorem \ref{Hauptsatz}}
\label{HS-bew}
The proof follows the usual lines developed in \cite{schaeffer:86} and in
\cite{baukun} for higher order approximations. In a first step the difference of the
fields, $\varphi-\varphi^D$ and the difference of its derivatives is estimated
in terms of the difference of the densities $h=f-f^D$. For that purpose we
shall use quite elaborated representations of the fields, which are  derived in the
appendix. Next we will calculate which Vlasov equation it is  that $h$
fulfills.  Using the
matching of the initial data and a Gronwall argument, it follows that
$|h|={\cal O}(c^{-4})$ which in turn  yields the announced estimates of the errors in the
fields.

As $\nabla\varphi$ enters the Vlasov equation with the factor $c^2$ we have to
be most precise in comparing $\nabla\varphi$ with
$\nabla\varphi^D$. Therefore, we give the reasoning for that term in some detail.  
In section  \ref{repapp-sect} below we will show that  the gradient of the  approximate field
$\varphi^D$ from (\ref{Darwin-def}) admits the following representation.
\begin{equation}\label{gphiD-rep}
   \nabla_x\varphi^D = \varphi^D_{x, \mathtt{ext}}
               +\varphi^D_{x, \mathtt{int}}
                +\varphi^D_{x, \mathtt{bd}}+{\cal O}_{{cpt}}(c^{-6})
\end{equation}
with
\begin{eqnarray}\label{gphiDext-rep}
   \varphi^D_{x, \mathtt{ext}}(t, x) & = & -\frac{1}{c^2}\intgr\int|z|^{-2}\bar{z}
   \Big(f^D-\frac{p^2}{2c^2}f_0\Big)(t, x+z, p)\,dp\,dz 
   \nonumber \\
   & & +\frac{1}{2c^4}\intgr\int\bar{z}\partial_t^2 f_0(t, x+z, p)\,dp\,dz,
   \\ \label{gphiDint-rep}
   \varphi^D_{x, \mathtt{int}} (t, x) & = & 
   -\frac{1}{c^2}\intkl\int|z|^{-2}\bar{z}f^D\tdxz\,dz
   \nonumber \\
   & & +\frac{1}{c^3}\intkl\int|z|^{-2}\big(2\bar{z}(\bar{z}\cdot p)-p\big)f^D\tdxzp\,dp\,dz
   \nonumber \\
   & & +\frac{1}{c^4}\intkl\int|z|^{-2}\big(-3\bar{z}(\bar{z}\cdot p)^2+(\bar{z}\cdot p) p+\frac{3}{2}\bar{z}p^2\big)
   f^D\tdxzp\,dp\,dz
   \nonumber \\
   & & -\frac{1}{c^4}\intkl\int|z|^{-1}\bar{z}\bar{z}\cdot \nabla_x\varphi_2 f^D\tdxzp\,dp\,dz
   \nonumber \\
   & & +\frac{1}{c^5}\intkl\int|z|^{-2}\Big(4(\bar{z}\cdot p)^3\bar{z}-(\bar{z}\cdot p)^2p-4(\bar{z}\cdot p)p^2\bar{z}
   +p^2p\Big)f^D\tdxzp\,dp\,dz
   \nonumber \\
   & & +\frac{1}{c^5}\intkl\int|z|^{-1}\bar{z}\Big(2(\bar{z}\cdot p)\bar{z}\cdot\nabla_x\varphi_2
   -p\cdot\nabla_x\varphi_2-\tilde{S}(\varphi_2)\Big)f^D\tdxzp\,dp\,dz 
   \nonumber \\
   & & -\frac{1}{3c^5}\intkl \partial_t\Big(\nabla_x\varphi_2 \mu_0\Big)(\hat{t}(z), x+z)\,dz, 
   \\ \label{gphiDbd-rep}
   \varphi^D_{x, \mathtt{bd}}(t, x) & = & \frac{1}{c^4t}\int_{|z|=ct}\int\bar{z}(\bar{z}\cdot p)\Big(
   1-\frac{(\bar{z}\cdot p)}{c}+\frac{(\bar{z}\cdot p)^2-p^2}{c^2}\Big) f^\circ(x+z,
   p)\,dp\,ds(z)
   \nonumber \\
   & & -\frac{t}{3c^4}\intgl\int\bar{z}(\bar{z}\cdot p)\partial_t^2f_0(0, x+z, p)\,dp\,ds(z)
   \nonumber \\
   & & -\frac{1}{3c^5}\intgl\int(\bar{z}\cdot p)p\dt f_0(0, x+z, p)\,dp\,ds(z)
   \nonumber
\end{eqnarray}   
where the subscripts `\texttt{ext}', `\texttt{int}' and `\texttt{bd}' refer to the exterior, interior
and boundary integration in $z$.  We also recall that $\bar{z}=|z|^{-1}z$ 
and $\hat{t}(z)=t-c^{-1}|z|$ is the retarded time. On the other hand, according to Section
\ref{repnord-sect} below, we have
\begin{equation}\label{gphi-rep}
    \nabla_x\varphi = \varphi_{x, \mathtt{ext}}
             +\varphi_{x, \mathtt{int}}
             +\varphi_{x, \mathtt{bd}}+{\cal O}_{{cpt}}(c^{-6})
\end{equation}
with  
\begin{eqnarray}\label{gphiext-rep} 
   \varphi_{x, \mathtt{ext}}(t, x) & = & -\frac{1}{c^2}\intgr\int|z|^{-2}\bar{z}
   \Big(f_0+t\dt f_0+\frac{t^2}{2}\dt^2 f_0+\frac{t^3}{6}\dt^3 f_0\Big)(0, x+z, p)\,dp\,dz 
   \nonumber \\
   & & +\frac{1}{2c^4}\intgr\int|z|^{-2}\bar{z}p^2(f_0+t\dt f_0)(0, x+z, p)\,dp\,dz
   \nonumber \\ 
   & & +\frac{1}{2c^4}\intgr\int\bar{z}\Big(\partial_t^2 f_0+t\dt^3 f_0\Big)(0, x+z, p)\,dp\,dz
   \nonumber \\
   & & -\frac{t}{c^4}\intgr\int|z|^{-2}\bar{z}\dt f_2(0, x+z, p)\,dp\,dz,
   \\ \label{gphiint-rep}
   \varphi_{x, \mathtt{int}} (t, x) & = & 
   -\frac{1}{c^2}\intkl\int|z|^{-2}\bar{z} f\tdxz\,dz
   \nonumber \\
   & & +\frac{1}{c^3}\intkl\int|z|^{-2}\big(2\bar{z}(\bar{z}\cdot p)-p\big)f\tdxzp\,dp\,dz
   \nonumber \\
   & & +\frac{1}{c^4}\intkl\int|z|^{-2}\big(-3\bar{z}(\bar{z}\cdot p)^2+(\bar{z}\cdot p) p+\frac{3}{2}\bar{z}p^2\big)
   f\tdxzp\,dp\,dz
   \nonumber \\
   & & -\frac{1}{c^2}\intkl\int|z|^{-1}\bar{z}\bar{z}\cdot \nabla_x\varphi f\tdxzp\,dp\,dz
   \nonumber \\
   & & +\frac{1}{c^5}\intkl\int|z|^{-2}\Big(4(\bar{z}\cdot p)^3\bar{z}-(\bar{z}\cdot p)^2p-4(\bar{z}\cdot p)p^2\bar{z}
   +p^2p\Big)f_0\tdxzp\,dp\,dz
   \nonumber \\
   & & +\frac{1}{c^3}\intkl\int|z|^{-1}\bar{z}\Big(2(\bar{z}\cdot p)\bar{z}\cdot\nabla_x\varphi
   -p\cdot\nabla_x\varphi-\tilde{S}(\varphi)\Big)f\tdxzp\,dp\,dz 
   \nonumber \\
   & & -\frac{1}{3c^5}\intkl\partial_t\Big(\nabla_x\varphi_2\mu_0\Big)(0, x+z)\,dz ,
   \\ \label{gphibd-rep}
   \varphi_{x, \mathtt{bd}}(t, x) & = & \varphi^D_{x, \mathtt{bd}}(t, x).
   \nonumber
\end{eqnarray}
 In order to verify (\ref{diff-esti}) we start by comparing the exterior
fields. Let $x\in B(0, R)$ with $R>0$ be fixed. Then we obtain from
(\ref{gphiDext-rep}) and (\ref{gphiext-rep}), taking into account that
\begin{equation}
   \int f_2(0, x,p)\,dp = 0
   \nonumber
\end{equation}
by (\ref{LVP}) and (\ref{IC-LVP}) as well as (\ref{SchrankeGeschwVP}), (\ref{SchrankeFeldVP}),
(\ref{TrLVP}) and (\ref{SchrankeLVP}),
\begin{eqnarray}
  \label{ext-est}
  \lefteqn{|\nabla_x\varphi_\mathtt{ext}(t, x)-\nabla_x\varphi^D_{\mathtt{ext}}(t,
  x)|}\hspace{1em}
  \nonumber \\
  & \leq & \frac{1}{c^2}\intgr|z|^{-2}\Big|\mu_0(t, x+z)-\big(\mu_0+t\dt\mu_0
  +\frac{t^2}{2}\dt^2\mu_0+\frac{t^3}{6}\dt^3\mu_0\big)(0, x+z)\Big|\,dz
  \nonumber \\
  & & +\frac{1}{2c^4}\intgr\Big|\dt^2\mu_0(t, x+z)-\big(\dt^2\mu_0+t\dt^3\mu_0\big)(0, x+z)\Big|\,dz
  \nonumber \\
  & & +\frac{1}{c^4}\intgr\int|z|^{-2}|f_2(t, x+z, p)-\big(f_2+t\dt f_2\big)(0, x+z, p)|\,dp\,dz
  \nonumber \\
  & & +\frac{1}{2c^4}\intgr\int|z|^{-2}p^2|f_0(t, x+z, p)-\big(f_0+t\dt f_0\big)(0, x+z, p)|\,dp\,dz
  \nonumber \\
  & \leq & \frac{M}{c^2}\intgr|z|^{-2}\bigg(\int_0^t(t-s)^3P_{VP}(s)^3K_{VP}(s)
  \mathbf{1}_{B(0, P_{VP}(s))}(x+z)\,ds\bigg)\,dz
  \nonumber \\
  & & +\frac{M}{c^4}\intgr\bigg(\int_0^t (t-s)P_{VP}(s)^3K_{VP}(s)
  \mathbf{1}_{B(0, P_{VP}(s))}(x+z)\,ds\bigg)\,dz
  \nonumber \\
  & & +\frac{M}{c^4}\intgr|z|^{-2}\bigg(\int_0^t (t-s)P_{VP}(s)^5K_{VP}(s)
  \mathbf{1}_{B(0, P_{VP}(s))}(x+z)\,ds\bigg)\,dz
  \nonumber \\
  & & +\frac{M}{c^4}\intgr|z|^{-2}\int_0^t(t-s)\bigg(P_{LVP}(s)^3K_{LVP}(s)
  \mathbf{1}_{B(0, P_{LVP}(s))}(x+z)\,ds\bigg)\,dz
  \nonumber \\
  & \leq & \frac{Mt^4}{c^2}\intgr|z|^{-2}\mathbf{1}_{B(0, R+M_0)}(z)\,dz
  +\frac{Mt^2}{c^4}\intgr\big(|z|^{-2}+1\big)\mathbf{1}_{B(0,
  R+M_0)}(z)\,dz
  \nonumber \\
  & \leq & M_Rc^{-6};  
\end{eqnarray}
note that here we have used 
\begin{equation}\label{SchrankeM0}
   M_0 = \max_{s\in [0, T]}\Big(P_{VP}(s)+K_{VP}(s)+P_{LVP}(s)+K_{LVP}(s)\Big)<\infty
   \nonumber
\end{equation}
and for instance
\begin{equation}
   \frac{t^4}{c^2}\intgr|z|^{-2}\mathbf{1}_{B(0,
   R+M_0)}(x+z)\,dz\leq \frac{t^4}{c^6t^4}\int_{|z|\leq R+M_0}|z|^2\,dz\leq M_R c^{-6}. 
   \nonumber
\end{equation}

To bound $|\varphi_{x, \mathtt{int}}(t, x)-\varphi^D_{x, \mathtt{int}}(t,
x)|$
we shall first treat 
the last term in (\ref{gphiDint-rep}) and (\ref{gphiint-rep}) respectively, 
which is  special in some way as it does not arise from the expansion of the integral kernels of $\nabla_x\varphi$.
Regarding this term, a crucial observation is that
by (\ref{VP}) and partial integration 
\begin{eqnarray}
   \label{Rad1}
   \lefteqn{\int\nabla_x\varphi_2\mu_0(t, x)\,dx  =  
   \frac{1}{4\pi}\bigg(\sum_{j=1}^3\int\partial_{x_i}\varphi_2\partial_{x_j}\partial_{x_j}\varphi_2(t, x)\,dx\bigg)_{i=1,2,3}}
   \nonumber \\
   & = & -\frac{1}{4\pi}\bigg(\sum_{j=1}^3\int\partial_{x_i}\partial_{x_j}\varphi_2\partial_{x_j}\varphi_2(t, x)\,dx\bigg)_{i=1,2,3}
   =  -\frac{1}{8\pi}\int\nabla_x|\nabla_x\varphi_2(t, x)|^2\,dx = 0
\end{eqnarray}
for all $t\in[0, \infty[$.
This yields
\begin{eqnarray}
   \label{Rad2}
   \lefteqn{-\frac{1}{3c^5}\intkl\dt \Big(\mu_0\nabla_x\varphi_2\Big)(\hat{t}(z),
    x+z)\,dz  =  -\frac{1}{3c^5}\intkl\dt\Big(\mu_0\nabla_x\varphi_2\Big)(t,
    x+z)\,dz+{\cal O}_{cpt}(c^{-6})}
    \nonumber \\
    & = & -\frac{\dt}{3c^5}\bigg(\intkl\Big(\mu_0\nabla_x\varphi_2\Big)(t,
    x+z)\,dz\bigg)
    +\frac{1}{3c^4}\intgl\Big(\mu_0\nabla_x\varphi_2\Big)(0, x+z)\,ds(z)+{\cal
    O}_{cpt}(c^{-6})
    \nonumber \\
    & = & \frac{\dt}{3c^5}\bigg(\intgr\Big(\mu_0\nabla_x\varphi_2\Big)(t,
    x+z)\,dz\bigg)
    +\frac{1}{3c^4}\intgl\Big(\mu_0\nabla_x\varphi_2\Big)(0, x+z)\,ds(z)+{\cal
    O}_{cpt}(c^{-6})
    \nonumber \\
    & = & \frac{1}{3c^5}\intgr\dt\Big(\mu_0\nabla_x\varphi_2\Big)(t, x+z)\,dz+{\cal
    O}_{cpt}(c^{-6})
\end{eqnarray}      
where we used  that for  $x\in B(0, R)$ 
\begin{eqnarray*}
   \lefteqn{\Big|\intkl\dt\Big(\mu_0\nabla_x\varphi_2\Big)(\hat{t}(z),
     x+z)-\dt\Big(\mu_0\nabla_x\varphi_2\Big)(t, x+z)\,dz\Big|}\hspace{5em}
   \nonumber \\
   & \leq &
   \frac{M}{c}\intkl|z|K_{V\!P}^2(t) P_{V\!P}^3(t)\mathbf{1}_{B(0,
   P_{V\!P})}(x+z)\,dz
   \nonumber \\
   &\leq & \frac{M}{c}\int|z|\mathbf{1}_{R+M_0}(z)\,dz\leq M_Rc^{-1}
\end{eqnarray*}
(note that $|\hat{t}(z)|\leq t$ for $|z|\leq ct$ and $K_{V\!P}$ and $P_{V\!P}$ are nondecreasing).
Again by (\ref{Rad1}) and same calculations we also have
\begin{equation}
  \label{Rad3}
  -\frac{1}{3c^5}\intkl\dt\Big(\mu_0\nabla_x\varphi_2\Big)(0, x+z)\, dz
  = \frac{1}{3c^5}\intgr\dt\Big(\mu_0\nabla_x\varphi_2\Big)(0, x+z)\, dz.
\end{equation}
Employing (\ref{Rad2}) and (\ref{Rad3}) we can estimate 
\begin{eqnarray}
   \label{Rad4}
   \lefteqn{\frac{1}{3c^5}\bigg|\intkl\dt\Big(\mu_0\nabla_x\varphi_2\Big)(\hat{t}(z), x+z)-
   \dt\Big(\mu_0\nabla_x\varphi_2\Big)(0, x+z)\,dz\bigg|}\hspace{3em}
   \nonumber \\
   & \leq & \frac{M}{c^5}\bigg|\intgr\dt\Big(\mu_0\nabla_x\varphi_2\Big)(t, x+z)-
   \dt\Big(\mu_0\nabla_x\varphi_2\Big)(0, x+z)\,dz\bigg|+M_Rc^{-6}
   \nonumber \\
   & \leq & \frac{M}{c^6}\intgr|z|K_{V\!P}^2(t)P_{V\!P}(t)^3\mathbf{1}_{B(0, P_{VP}(t) }(x+z)\,dz+M_Rc^{-6}\leq M_Rc^{-6}
\end{eqnarray}  
for all $t\in[0, T]$ and $x\in B(0, R) $.  
  
Next we recall from 
\cite[Thm.3]{calee} that
\begin{equation}
  \label{new-app}
  |\nabla_x\varphi(t, x)-c^{-2}\nabla_x\varphi_2(t, x)| =  \mathcal{O}(c^{-3});
\end{equation}
actually in \cite{calee} the initial conditions  are different, but we only
added terms of order $c^{-4}$ so that an inspection of the proof in
\cite{calee} leads to (\ref{new-app}).
Furthermore, in section \ref{repapp-sect}, (\ref{dtphiD-rep}) and section \ref{repnord-sect},
(\ref{dtphi-est})  it is shown that
\begin{equation} 
   \label{new-app2}
   |\dt\varphi^D(t, x)|+|\dt\varphi(t, x)|   =   {\cal O}(c^{-2}).
\end{equation}
Also, we define 
\begin{equation}
   H(t) = \sup\{|f(s, x, p)-f_0(s, x, p)-c^{-2}f_2(s, x,
   p)|\,:\,x\in\R^3,\,p\in\R^3,\,s\in[0, t]\}.
\end{equation}
Now we have to proceed in two steps: Using (\ref{new-app}) and (\ref{new-app2})  prove
\begin{eqnarray}
   \label{darwin1-est}
   |\nabla_x\varphi^D(t, x)-\nabla_x\varphi(t, x)| & \leq & \frac{1}{c^2}M_R(c^{-3}+H(t))
   \nonumber \\ 
   |\dt\varphi^D(t, x)-\dt\varphi(t, x)| & \leq  &  M(c^{-3}+H(t)),
   \nonumber 
\end{eqnarray} 
employ this to derive
\begin{equation}
   H(t)\leq Mc^{-3}
   \nonumber
\end{equation}
for all $t\in[0, T]$, which in turn gives
\begin{eqnarray}
   \label{darwin2-est}
   |\nabla_x\varphi^D(t, x)-\nabla_x\varphi(t, x)| & \leq & M_Rc^{-5}
   \nonumber \\ 
   |\dt\varphi^D(t, x)-\dt\varphi(t, x)| & \leq  &  Mc^{-3}
   \nonumber 
\end{eqnarray}  
and especially as $c^{-4}\nabla_x\varphi_2={\cal O}(c^{-4})$
\begin{equation}
   \label{darwin3-est}
   |c^{-2}\nabla_x\varphi_0(t, x)-\nabla_x\varphi(t, x)| \leq  M_Rc^{-4}.
\end{equation}
In the second step this procedure will be repeated where   estimate
(\ref{new-app}) is replaced by (\ref{darwin3-est}),
which will give the announced estimates. To avoid redundancies we shall only carry out the second step, and therefore assume 
that (\ref{darwin3-est}) is  already proved. 

Introducing the constant
\begin{equation}
   M_1 = \max_{s\in[0, T]}\Big(P_{VN}(s)+P_{VP}(s)+P_{LVP}(s)\big)<\infty
   \nonumber
\end{equation}
it follows that
$f(s, x, p)=f_0(s, x, p)=f_2(s, x, p)=0$ for $x\in\R^3$, $|p|> M_1$
and $s\in[0, T]$.  Also if $R_0>0$ is chosen such that  $f^\circ(x, p)=0$ for
$|x|\geq R_0$, defining the constant 
\begin{equation}
   M_2 = R_0+TM_1+\max_{s\in[0, T]}\Big(P_{VP}(s)+P_{LVP}(s)\Big)<\infty,
\end{equation}
we have  that $f(s, x, p)=f_0(s, x, p)=f_2(s, x, p)=0$ for $|x|\geq M_2$,
$p\in\R^3$ and $s\in [0, T]$. Let again $x\in B(0, R)$ with $R>0$ be fixed. From
(\ref{gphiint-rep}), (\ref{gphiDint-rep}), (\ref{Rad4}), (\ref{SchrankeM0}), (\ref{new-app}), (\ref{darwin3-est})
(\ref{IC-matched}) and $0\le\hat{t}(z)\leq t$ for $|z|\leq ct$ we obtain
\begin{eqnarray}
   \label{int-est}
   \lefteqn{|\varphi_{\mathtt{int}}(t, x)-\varphi^D_{\mathtt{int}}(t,
   x)|}\hspace{1em}
   \nonumber \\
   & \leq &
   \frac{1}{c^2}\intkl\int|z|^{-2}\Big|(f-f^D)\tdxzp\Big|\,dp\,dz
   \nonumber \\
    & & +\frac{1}{c^3}\intkl\int|z|^{-2}\big|2\bar{z}(\bar{z}\cdot p)-p\big|\big|(f-f^D)\tdxzp\Big|\,dp\,dz
   \nonumber \\
   & & +\frac{1}{c^4}\intkl\int|z|^{-2}\big|-3\bar{z}(\bar{z}\cdot p)^2+(\bar{z}\cdot p) p+\frac{3}{2}\bar{z}p^2\big|
   \Big|(f-f^D)\tdxzp\Big|\,dp\,dz
   \nonumber \\
   & & +\frac{1}{c^2}\intkl\int|z|^{-1}\Big| (\nabla_x\varphi-\frac{1}{c^2}\nabla_x\varphi_0\Big| f\tdxzp\,dp\,dz
   \nonumber \\
   & & +\frac{1}{c^4}\intkl\int|z|^{-1}\Big|\nabla_x\varphi_2(f-f^D)\tdxzp)\Big|\,dp\,dz
   \nonumber \\
   & & +\frac{1}{c^5}\intkl\int|z|^{-2}\Big|4(\bar{z}\cdot p)^3\bar{z}+(\bar{z}\cdot p)^2p-4(\bar{z}\cdot p)p^2\bar{z}
   +\frac{3p^2}{2}p\Big|\Big|(f-f^D)\tdxzp\Big|\,dp\,dz
   \nonumber \\
   & & +\frac{M}{c^3}\intkl\int|z|^{-1}\bigg(|p|\Big|\nabla_x\varphi-\frac{1}{c^2}\nabla_x\varphi_2\Big|
   +\Big|\tilde{S}(\varphi)-\frac{1}{c^2}\tilde{S}(\varphi_2)\Big|\bigg)f\tdxzp\,dp\,dz 
   \nonumber \\
   & & +\frac{M}{c^5}\intkl\int|z|^{-1}\Big(|p||\nabla_x\varphi_2|+|\tilde{S}(\varphi_2)|\Big)\Big|f-f^D\Big|\tdxzp\,dp\,dz+M_Rc^{-6}
   \nonumber \\
   & \leq & \frac{M}{c^2}(M_1^3+M_1^6)H(t)\intkl|z|^{-2}\mathbf{1}_{B(0, M_2)}(x+z)\,dz
   \nonumber \\
   & & +\frac{M}{c^4}(M_1^3+M_1^4)K_{VP}(T)H(t)\intkl|z|^{-1}\mathbf{1}_{B(0, M_2)}(x+z)\,dz
   \nonumber \\
   & & +\frac{M_R}{c^6}(M_1^4+M_1^4)||f^\circ||_{x,p}\intkl\mathbf{1}_{B(0, M_2)}(x+z)\,dz+\frac{M_R}{c^6}
   \nonumber \\
   &\leq & c^{-2}M_R(H(t)+c^{-4})
\end{eqnarray}
since for instance 
\begin{equation}
   \intkl|z|^{-1}\mathbf{1}_{B(0,
   M_2)}(x+z)\,dz\leq\int_{|z|\leq R+M_2}|z|^{-1}\,dz\leq M_R.
   \nonumber
\end{equation}
Recalling that $\varphi_{x, \mathtt{bd}}=\varphi^D_{x, \mathtt{bd}}$, we can
summarize (\ref{gphiD-rep}), (\ref{gphi-rep}), (\ref{ext-est}) and
(\ref{int-est}) as
\begin{equation}
   \label{gphidarwin-est}
   c^2|\nabla_x\varphi(t, x)-\nabla_x\varphi^D(t, x)|\leq M_R(H(t)+c^{-4})
\end{equation}
for $x\in B(0, R)$ and $t\in[0, T]$. 
Formulas (\ref{phiD-rep}),(\ref{phi-approx}), (\ref{dtphiD-rep}), (\ref{dtphi-approx}) and an analog calculation also
lead to
\begin{eqnarray}
  \label{phidarwin-est}
  |\varphi(t, x)-\varphi^D(t, x)| &\leq& M(H(t)+c^{-4})
  \\ \label{dtphidarwin-est}
  |\dt\varphi(t, x)-\dt\varphi^D(t, x)| &\leq&  M(H(t)+c^{-4})
\end{eqnarray}
for $t\in [0, T]$. It remains to estimate $h = f-f^D$. Using (\ref{VN}),
(\ref{Darwin-def}), (\ref{VP}) and (\ref{LVP}), it is found that
\begin{eqnarray}
   \lefteqn{S(h)-\Big[S(\varphi)p+\gamma c^2\nabla_x\varphi\Big]\cdot\nabla_p h}
   \hspace{0.5em}
   \nonumber \\
   & = & - S(f^D)+\Big[S(\varphi)p+\gamma c^2\nabla_x\varphi\Big]\cdot\nabla_p f^D+4S(\varphi)f
   \nonumber \\
   &  = & -\Big[\hat{p}-p+\frac{1}{2c^2}p^2p\Big]\cdot\nabla_x f_0-\frac{1}{c^2}(\hat{p}-p)\cdot\nabla_x f_2
   +\Big[S(\varphi)-\frac{1}{c^{2}}\tilde{S}(\varphi_2)\Big]p\cdot\nabla_p f_0
   \nonumber \\
   & & +\frac{1}{c^2}S(\varphi)p\cdot\nabla_p f_2+\Big[\gamma c^2\nabla_x\varphi
    -\Big(1-\frac{p^2}{2c^2}\Big)\nabla_x\varphi_2-\frac{1}{c^2}\nabla_x\varphi_4\Big]\cdot\nabla_p f_0
   \nonumber \\
   & &
   +\Big[\gamma\nabla_x\varphi-\frac{1}{c^2}\nabla_x\varphi_2\Big]\cdot\nabla_p
   f_2+4S(\varphi)h-4\Big(\frac{1}{c^2}\tilde{S}(\varphi_2)-S(\varphi)\Big)f_0+\frac{4}{c^2}S(\varphi)f_2.
   \nonumber
\end{eqnarray}
If $|p|\leq M_1$, then also $|\hat{p}|=(1+c^{-2}p^2)^{-1/2}|p|\leq|p|\leq M_1$ uniformly in $c$, and hence
\begin{equation}
   \frac{1}{c^2}\Big|\hat{p}-p\Big|+\Big|\hat{p}-p+\frac{1}{2c^2}p^2p\Big|\leq M c^{-4}.
   \nonumber
\end{equation}
In view of the bounds (\ref{SchrankeGeschwVP}), (\ref{SchrankeLVP}) and (\ref{SchrankeFelder}), $S(\varphi)={\cal O}(c^{-2})$ as shown in 
section \ref{repnord-sect}, (\ref{gphi-est}) and (\ref{dtphi-est}), 
thus by (\ref{gphidarwin-est}) and (\ref{dtphidarwin-est})
\begin{gather}
   \Big|S(h)(t, x, p)-\Big[S(\varphi)(t, x, p)+\gamma c^2\nabla_x\varphi(t, x)\Big]
   \cdot\nabla_p f(t, x, p)\Big|
   \nonumber\\
    \leq  M\big(c^{-4}+H(t)\big)\label{h-est}
\end{gather}
for all $|x|\leq M_2$, $|p|\leq M_1$ and $t\in[0, T]$. But in $\{(t, x, p)\,:\,|x|>M_2\}\cup\{(t, x, p)\,:\,|p|>M_1\}$ 
we have $h=f-f^D=0$ by the above definition of  $M_1$ and $M_2$. Accordingly, (\ref{h-est}) 
is satisfied for all $x\in\R^3,\,p\in\R^3$ and $t\in[0, T]$. Since $h(0, x, p)=0$ the argument from \cite{calee} (see also 
\cite[p.416]{schaeffer:86}) yields
\begin{equation}
   H(t)\leq\int_0^t M(c^{-4}+H(s))ds, 
   \nonumber
\end{equation}
and therefore $H(t)\leq M c^{-4}$ for $t\in[0, T]$. Then, due to (\ref{phidarwin-est})--(\ref{dtphidarwin-est})
$c^2|\nabla_x\varphi(t, x)-\nabla_x\varphi^D(t, x)|\leq M_R c^{-4}$ for 
$|x|\leq R$ and $t\in[0, T]$ as well as $|\varphi(t, x)-\varphi^D(t, x)|
+|\dt\varphi(t, x)-\dt\varphi^D(t, x)|\leq M c^{-4}$ for $x\in\R^3$ and $t\in[0,T]$.
This completes the proof of Theorem \ref{Hauptsatz}.
{\hfill$\Box$}\bigskip 

\setcounter{equation}{0}

\section{Proof of Theorem \ref{DVN-thm}}
\label{DVN-bew}

In this section we will be sketchy and omit many details,
since the proof is more or less a repetition of what has been said before.
First let us assume that there is a $C^2$-solution $(f^\ast, \psi^\ast, E^\ast)$
of (\ref{DVNc}) existing on a time interval $[0, T^\ast]$ for some $T^\ast>0$
such that ${\rm supp}\,f^\ast(t, \cdot, \cdot)\subset\R^3\times\R^3$ is compact
for all $t\in [0, T^\ast]$. Then, after partial integration
\begin{eqnarray}
  \label{psiast-def}
  \psi^\ast(t, x) & = & \frac{1}{c^2}\int\int|z|^{-2}\bar{z}\cdot p f^\ast(t,
  x+p, p)\,dp\,dz
  \\ \label{East-def}
  \Delta E^\ast (t, x) & = & \frac{4\pi}{c^2}\big(\nabla\mu^\ast+\dt\nabla\psi^\ast\big).
\end{eqnarray}
Since $f^\ast(0, x, p)=f^\circ(x, p)$, it follows that $\psi^\ast(0, \cdot)$
is determined by $f^\circ$.

In order to compute the Poisson integral for $E^\ast$, we calculate by means
of the transformation $y=w-z,\;dy=dw,$ and using (\ref{formelv}) below,
\begin{eqnarray*}
  \lefteqn{\frac{1}{c^2}\Big(\Delta^{-1}\dt\nabla\psi^\ast\Big)(t, x)}
  \\
  & = & -\frac{1}{c^2 4\pi}\int|x-y|^{-1}\nabla\dt\psi^\ast(t, y)\,dy
  \\
  & = & -\frac{1}{c^4 4\pi}\int|x-y|^{-1}\nabla_y\bigg(\int\int|z|^{-2}(\bar{z}\cdot
  p)\dt f^\ast(t, y+z, p)\,dp\,dz\bigg)\,dy
  \\
  & = & -\frac{1}{c^4
  4\pi}\bigg[\int\bigg(\int|z|^{-2}\bar{z}|x-w-z|^{-1}\,dz\bigg)\cdot\int
  p\partial_{w_i}\dt f^\ast(t, w, p)\,dp\,dw\bigg]_{i=1,2,3}
  \\
  & = & -\frac{1}{2c^4
  }\int\int|w-x|^{-1}(w-x)\cdot p\nabla_w \dt f^\ast(t, w, p)\,dp\,dw
  \\
  & = & -\frac{1}{2c^4
  }\int\int(\bar{z}\cdot p)\nabla_z\dt f^\ast(t, x+z, p)\,dp\,dz
  \\
  & = & \frac{1}{2c^4
  }\int\int|z|^{-1}\big(-\bar{z}(\bar{z}\cdot p)+p\big)\dt f^\ast(t, x+z, p)\,dp\,dz.
\end{eqnarray*}
If we invoke the Vlasov equation for $f^\ast$ and integrate by parts, this can
be rewritten as
\begin{eqnarray*}
   \lefteqn{\frac{1}{c^2}\Big(\Delta^{-1}\dt\nabla\psi^\ast\Big)(t, x)}
  \\
  & = & \frac{1}{2c^4
  }\int\int|z|^{-1}\bigg(-\bar{z}(\bar{z}\cdot
  p)+p\big)\Big(-(1-\frac{p^2}{2c^2})p\cdot \nabla_x
  f^\ast
  \\
  & & +\Big[(\psi^\ast+p\cdot E^\ast)p+c^2\Big(1-\frac{p^2}{2c^2}\Big)E^\ast\Big]\cdot \nabla_p f^\ast
  +4(\psi^\ast+p\cdot E^\ast)f^\ast\bigg)(t, x+z, p)\,dp\,dz
  \\
  & = & \frac{1}{2c^4}\int\int|z|^{-2}\Big(-2(\bar{z}\cdot p)p+3(\bar{z}\cdot
  p)^2\bar{z}-p^2\bar{z}\Big)\Big(1-\frac{p^2}{2c^2}\Big)f^\ast(t, x+z,
  p)\,dp\,dz
  \\
  & & -\frac{1}{2c^4}\int\int|z|^{-1}\bigg\{(1-\bar{z}\otimes\bar{z})
  \Big[(\psi^\ast+p\cdot E^\ast)p+c^2\Big(1-\frac{p^2}{2c^2}\Big)E^\ast\Big]f^\ast
  \\
  & & \hspace{7.5em}+\big(\bar{z}(\bar{z}\cdot
  p)-p\big)\big(\psi^\ast+p\cdot E^\ast\big)f^\ast\bigg\}(t, x+z, p)\,dp\,dz.
\end{eqnarray*}
Therefore the solution $E^\ast$ of (\ref{East-def}) has the representation 
\begin{eqnarray}
  \label{East-rep}
  E^\ast(t, x) & = &
  \Delta^{-1}\Big(\frac{4\pi}{c^2}\nabla\mu^\ast\Big)+\Delta^{-1}\Big(\frac{1}{c^2}\nabla\dt\psi^\ast\Big)
  \nonumber \\
  & = & -\frac{1}{c^2}\int|z|^{-2}\bar{z}\mu^\ast(t, x+z)\,dz
  \nonumber \\
  & & +\frac{1}{2c^4}\int\int|z|^{-2}\Big(-2(\bar{z}\cdot p)p+3(\bar{z}\cdot
  p)^2\bar{z}-p^2\bar{z}\Big)\Big(1-\frac{p^2}{2c^2}\Big)f^\ast(t, x+z,
  p)\,dp\,dz
  \nonumber \\
  & & -\frac{1}{2c^4}\int\int|z|^{-1}\bigg\{(1-\bar{z}\otimes\bar{z})
  \Big[(\psi^\ast+p\cdot E^\ast)p+c^2\Big(1-\frac{p^2}{2c^2}\Big)E^\ast\Big]f^\ast
  \nonumber \\
  & & \hspace{7.5em}+\big(\bar{z}(\bar{z}\cdot
  p)-p\big)\big(\psi^\ast+p\cdot E^\ast\big)f^\ast\bigg\}(t, x+z, p)\,dp\,dz.
\end{eqnarray}
In particular, if we evaluate this relation at $t=0$ the Banach fix point
theorem applied  in $C_b(\R^3)$ shows that for $c\ge c^\ast$ sufficiently
large the function $E^\ast(0, \cdot)$ is uniquely determined by
$f^\circ=f^\ast(0, \cdot, \cdot)$. Thus $f^\circ$ alone fixes $\psi^\circ$ and
$E^\circ$.

Concerning the local and uniform (in $c$) existence of a solution to
(\ref{DVNc}) one can use (\ref{psiast-def}) and (\ref{East-rep}) to follow the
usual method by setting up an iteration scheme for which convergence can be
verified on a small time interval: cf. \cite[Sect. 5.8]{glassey:96} and also
\cite{carein}.
For comparison of $E^\ast$ with $\nabla_x\varphi^D$ we give an alternative
expression of $\nabla_x\varphi^D$.
Using (\ref{VP}) and partial integration twice  we have
\begin{eqnarray}
  \label{phifour-alt}
  \lefteqn{-\nabla_x\frac{1}{2c^4}\int|z|\dt^2\mu_0(t, x+z)\,dz  =  
  \frac{1}{2c^4}\int\bar{z}\dt^2\mu_0(t, x+z)\,dz}\hspace{2em}
  \nonumber \\
  & = & \frac{1}{2c^4}\int\int\bar{z}\dt\Big(-p\cdot\nabla_x f_0+\nabla_x\varphi_2\cdot
  \nabla_pf_0\Big)(t, x+z, p)\,dp\,dz
  \nonumber \\
  & = & \frac{1}{2c^4}\int\int|z|^{-1}\Big(-(\bar{z}\cdot p)\bar{z}+p\Big)\dt
  f_0(t, x+z, p)\,dp\,dz
  \nonumber \\
  & = & \frac{1}{2c^4}\int\int|z|^{-1}\Big(-(\bar{z}\cdot p)\bar{z}+p\Big)
  \Big(-p\cdot\nabla_xf_0+\nabla_x\varphi_2\cdot\nabla_p f_0\Big)(t, x+z, p)\,dp\,dz
  \nonumber \\
  & = & \frac{1}{2c^4}\int\int|z|^{-2}\Big(-2(\bar{z}\cdot p)p+3(\bar{z}\cdot
  p)^2\bar{z}-p^2\bar{z}\Big)f_0(t, x+z, p)\,dp\,dz
  \nonumber \\
  & & -\frac{1}{2c^4}\int|z|^{-1}(1-\bar{z}\otimes\bar{z})\nabla_x\varphi_2\mu_0(t, x+z)\,dz.
\end{eqnarray}
Together (\ref{gphiD-def}), (\ref{phifour-alt}) (\ref{East-rep}) and
(\ref{dtphiD-rep}), (\ref{psiast-def}) reveals
the analogy of $E^\ast$ to $\nabla_x\varphi^D$ and $\psi^\ast$ to
$\dt\varphi^D$ respectively at the relevant orders of $c^{-1}$. Comparison of
(\ref{phiast-def}) with (\ref{VP}), (\ref{LVP}) and (\ref{Darwin-def}) gives the analogy of the initial
values of (\ref{VN}). Finally, by similar arguments as used in the proof of
Theorem \ref{Hauptsatz} it can be shown that solutions of (\ref{DVNc})
approximate solutions of (\ref{VN}) up to an error of order $c^{-4}$.
 

\setcounter{equation}{0}

\section{Appendix}
\label{append}

\subsection{Representation Formulas}

\subsubsection{Representation of the approximation force field}
\label{repapp-sect}
Here we will present representation formulas for the approximate field and its
derivatives. As the calculations are  lengthy and arguments are  similar we shall only
sketch the computations leading to the formula
for $\nabla_x\varphi^D$  and merely give the formulas for $\varphi^D$ and $\dt\varphi^D$. 

We recall $ \varphi^D =  c^{-2}\varphi_2 + c^{-4} \varphi_4 $ where
\begin{eqnarray*}
   \varphi_2(t, x) & = & -\int |z|^{-1}\mu_0(t, x+z)\,dz \\
   \varphi_4(t, x) & = & -\frac{1}{2}\int |z|\partial_t^2\mu_0(t, x+z)\,dz
         -\int |z|^{-1}\mu_2(t, x+z)\,dz,
\end{eqnarray*} 
thus, using $\mu_0=\int f_0(\cdot ,\cdot, p)\,dp$ and $\mu_2=\int
\Big(f_2-\frac{p^2}{2}f_0\Big)(\cdot , \cdot , p)\,dp$
\begin{eqnarray}
   \label{gphiD-def}
   \nabla_x\varphi^D(t, x) & = & 
   -\frac{1}{c^2}\int\int|z|^{-2}\bar{z}\Big(f^D-\frac{p^2}{2c^2}f_0\Big)(t, x+z, p)\,dp\,dz
   \nonumber \\
   & &  +\frac{1}{2c^4}\int\bar{z}\dt^2\mu_0(t, x+z)\,dz.
\end{eqnarray} 
We split the domain of integration in $\{|z|>ct|\}$ and $\{|z|<ct\}$; note that the exterior part gives 
$\varphi_{x, \mathtt{ext}}$. 
To handle the interior part $\{|z|<ct\}$ we expand the densities
w.r.t. $t$ about the retarded time 
\begin{equation*}
   \hat t (z) := t - c^{-1} |z|.
\end{equation*}
To begin with we have
\begin{eqnarray}\label{Entwicklung-ret}
   \lefteqn{-\frac{1}{c^2}\intkl|z|^{-2}\bar{z}\mu_0(t, x+z)\,dz}  
   \nonumber \\
   & = & -\frac{1}{c^2}\intkl|z|^{-2}\bar{z}\Big(1+\frac{|z|}{c}\dt+\frac{|z|^2}{2c^2}\dt^2+\frac{|z|^3}{6c^3}\dt^3\Big)\mu_0\tdxz\,dz
   \nonumber \\ 
   & &-\frac{1}{6c^2}\intkl|z|^{-2}\bar{z}\int_{\hat{t}(z)}^t(t-s)^3\dt^4\mu_0(s, x+z)\,ds\,dz
\end{eqnarray} 
Since $\partial_t\mu_0+\int p \cdot \nabla_x\, f_0 \,dp=0$ by (\ref{VP}), we also find
\begin{eqnarray}\label{phiI2}
   \lefteqn{-\frac{1}{c^3}\int_{|z|\leq
   ct}|z|^{-1}\bar{z}\partial_t\mu_0(\hat{t}(z), x+z)\,dz
   =\frac{1}{c^3}\int_{|z|\leq ct}|z|^{-1}\bar{z}\Big(\int p \cdot \nabla_x 
   f_0(\hat{t}(z), x+z, p)\,dp\,dz}\hspace{2em}
   \nonumber \\
   & = & \frac{1}{c^3}\int_{|z|\leq ct}\int|z|^{-1}\bar{z} p\cdot
   \Big(\nabla_z [f_0(\hat{t}(z), x+z, p)]
   +c^{-1}\bar{z}\,\partial_t f_0(\hat{t}(z), x+z, p)\Big)\,dp\,dz
   \nonumber \\
   & = & \frac{1}{c^4t}\int_{|z|=ct}\int\bar{z}(\bar{z}\cdot p) f^\circ(x+z,
   p)\,dp\,ds(z)
   \nonumber \\
   & & -\frac{1}{c^3}\intkl\int|z|^{-2}\Big(-2\bar{z}(\bar{z}\cdot p)+p\Big)f_0\tdxzp\,dp\,dz
   \nonumber \\
   & & +\frac{1}{c^4}\int_{|z|\leq ct}\int|z|^{-1} \bar{z}(\bar{z}\cdot p)
   \,\partial_t f_0(\hat{t}(z), x+z, p)\,dp\,dz,
\end{eqnarray}
observe that $\hat{t}(z)=0$ for $|z|=ct$ was used for the boundary term. 
Using (\ref{SchrankeGeschwVP}), (\ref{SchrankeFeldVP}) and (\ref{SchrankeLVP}) the
last term in (\ref{phiI2}) as well as the remaining terms coming from the interior part of
(\ref{gphiD-def}) and from (\ref{Entwicklung-ret}) are  in ${\cal O}_{cpt}(c^{-4})$;
note that $|x|<R$ for some $R>0$ together with the support properties of $f_0$
imply that we only have to integrate in $z$ over a set which is uniformly
bounded in $c\geq 1$.
Thus the leading terms in the orders $c^{-2}$ and $c^{-3}$ are specified.  
The following scheme emerges,
  \begin{itemize}
      \item expand the several terms in the interior part of (\ref{gphiD-def}) up
            to order $c^{-5}$, the error term is in ${\cal O}_{cpt}(c^{-6})$;        
      \item replace  time derivatives on the densities $f_0$ and $f_2$ via Vlasov's equations  by
            space derivatives according to (\ref{VP}) and (\ref{LVP});
      \item carry out  partial integrations, using
            \[
                (\nabla_x g)\tdxz=\nabla_z[g\tdxz]+c^{-1}\bar{z}\dt g \tdxz
            \]
            in the case of derivatives with respect to $x$.
           
  \end{itemize}       
Following this scheme one can calculate one after another the leading terms up to order $c^{-5}$. 

Addition of some additional terms in ${\cal O}(c^{-6})$, as
e.g. $\frac{1}{2c^6}\intkl\int |z|^{-2}\bar{z}p^2 f_2\tdxzp\,dp\,dz$, then gives (\ref{gphiDint-rep}) and (\ref{gphiDbd-rep}).  

One can also calculate 
\begin{equation}\label{phiD-rep}
   \varphi^D = \varphi^D_{\mathtt{ext}}+\varphi^D_{\mathtt{int}}+\varphi^D_{\mathtt{bd}}+
   {\cal O}(c^{-4})
\end{equation}
with
\begin{eqnarray*}\label{phiDext-rep}
   \varphi^D_{\mathtt{ext}}(t,x) & = & 
   - \frac{1}{c^2}\int_{|z|>ct}|z|^{-1}\big(\mu_0+c^{-2}\mu_2\big)(x+z, t)\,dz
   \nonumber \\
   & & -\frac{1}{2c^4}\int_{|z|\geq ct}|z|\partial_t^2\mu_0(t, x+z)\,dz, 
   \\ \label{phiDint-rep}
   \varphi^D_{\mathtt{int}}(t, x) & = &
   -\frac{1}{c^2}\int_{|z|\leq 
   ct}\int|z|^{-1}f^D(\hat{t}(z), x+z, p)\,dp\,dz, 
   \\ \label{phiDbd-rep}
   \varphi^D_{\mathtt{bd}} (t, x) & = &   
   +\frac{1}{c^3}\int_{|z|=ct}\int (\bar{z}\cdot p)\,f^\circ(x+z, p)\,dp\,ds(z)
\end{eqnarray*} and 
\begin{equation}
   \label{dtphiD-rep}
   \dt\varphi^D =
   \varphi^D_{t,\mathtt{ext}}+\varphi^D_{t, \mathtt{int}}+\varphi^D_{t, \mathtt{bd}}
   +{\cal O}(c^{-4})
\end{equation}
is derived in the same manner with
\begin{eqnarray*}
   \label{dtphiDext-rep}
   \varphi^D_{t,\mathtt{ext}}(t, x) & = & 
   \frac{1}{c^2}\intgr\int|z|^{-2}(\bar{z}\cdot
   p)f^D(t, x+z, p)\,dp\,dz,
   \\ \label{dtphiDint-rep}
   \varphi^D_{t, \mathtt{int}}(t, x) & = &  
   \frac{1}{c^2}\intkl\int|z|^{-2}(\bar{z}\cdot p)f^D\tdxzp\,dp\,dz
   \nonumber \\
   & &
   -\frac{1}{c^3}\intkl\int|z|^{-1}\bar{z}\cdot\nabla_x\varphi_2f^D\tdxzp\,dp\,dz
   \nonumber \\
   & &+\frac{1}{c^3}\intkl\int|z|^{-2}\Big(p^2-2(\bar{z}\cdot p)^2\Big)f^D\tdxzp,
   \\ \label{dtphiDbd-rep}
   \varphi^D_{t, \mathtt{bd}}(t, x) & = & 
   -\frac{1}{c^4t}\intgl\int(\bar{z}\cdot p)^2f_0(0, x+z, p)\,dp\,ds(z).
\end{eqnarray*}

\subsubsection{Representation of the Nordstr{\"o}m force-field}
\label{repnord-sect}
We recall the following representation from \cite[Proposition 3; Proposition 4]{calee}
\begin{eqnarray}\label{phi}
   \varphi & = & \varphi_D+\varphi_S
   \\ \label{dtphi}
   \partial_t\varphi & = & \varphi_{t, D}+\varphi_{t,
   BD}+\varphi_{t,a}+\varphi_{t,b}+\varphi_{t,c} 
   \\ \label{gphi}
   \nabla_x\varphi &=& \varphi_{x, D}+\varphi_{x,
   BD}+\varphi_{x,a}+\varphi_{x,b}+\varphi_{x,c}
\end{eqnarray}
where
\begin{eqnarray*}
   \varphi_D(t, x) & = &  \partial_t\bigg(\frac{t}{4\pi}\int_{|\omega|=1}
                          \varphi^0(x+ct\omega)\,d\omega\bigg)
                          +\frac{t}{4\pi}\int_{|\omega|=1}\varphi^1(x+ct\omega)\,d\omega \\
   \varphi_S(t, x) & = & -\frac{1}{c^2}\int_{|z|\leq
                          ct}\int|z|^{-1}\gamma(p)f(\hat{t}(z), x+z, p)\,dp\,dz \\
   \varphi_{t, D}(t, x)  & = & \partial_t\varphi_D(t, x)  \\
   \varphi_{t, BD}(t, x) & = & -c^{-1}(ct)^{-1} \int_{|z|=ct}\int bd^{\varphi_t}
                          f^\circ(x+z,p)\,dp\,ds(z) \\
   \varphi_{t, a}(t, x)  & = & -c^{-2}\int_{|z|\leq ct}\int
                          a^{\varphi_t}f(\hat{t}(z), x+z, p)\,dp\,\frac{dz}{|z|^2} \\
   \varphi_{t, b}(t, x)  & = & -c^{-2}\int_{|z|\leq ct}\int
                          b^{\varphi_t}S(\varphi)f(\hat{t}(z), x+z, p)\,dp\,\frac{dz}{|z|} \\ 
   \varphi_{t, c}(t, x)  & = & -c^{-1}\int_{|z|\leq ct}\int
                          c^{\varphi_t}(\nabla_x\varphi)f(\hat{t}(z), x+z, p)\,dp\,\frac{dz}{|z|}    
\end{eqnarray*}
and
\begin{eqnarray*}
   \varphi_{x, D}(t, x)  & = & \nabla_x\varphi_D(t, x) \\
   \varphi_{x, BD}(t, x) & = & -c^{-2}(ct)^{-1} \int_{|z|=ct}\int bd^{\varphi_x} 
                          f^\circ(x+z,p)\,dp\,ds(z) \\
   \varphi_{x, a}(t, x)  & = & -c^{-2}\int_{|z|\leq ct}\int
                          a^{\varphi_x}f(\hat{t}(z), x+z, p)\,dp\,\frac{dz}{|z|^2} \\
   \varphi_{x, b}(t, x)  & = & -c^{-3}\int_{|z|\leq ct}\int
                          b^{\varphi_x}S(\varphi)f(\hat{t}(z), x+z, p)\,dp\,\frac{dz}{|z|} \\ 
   \varphi_{x, c}(t, x)  & = & -c^{-2}\int_{|z|\leq ct}\int
                          c^{\varphi_x}(\nabla_x\varphi)f(\hat{t}(z), x+z, p)\,dp\,\frac{dz}{|z|} 
\end{eqnarray*}
where the kernels are
\begin{eqnarray*}
   bd^{\varphi_t} & = & \gamma(p)(1+c^{-1}\bar{z}\cdot\hat{p})^{-1} \\
   a^{\varphi_t}  & = & -\gamma (1+c^{-1}\bar{z}\cdot\hat{p})^{-2}
                        \hat{p}\cdot(\bar{z}+c^{-1}\hat{p})        \\ 
   b^{\varphi_t}  & = & \gamma (1+c^{-1}\bar{z}\cdot\hat{p})^{-2}
                        (\bar{z}+c^{-1}\hat{p})^2        \\
   c^{\varphi_t}  & = & \gamma^3 (1+c^{-1}\bar{z}\cdot\hat{p})^{-2}
                        (\bar{z}+c^{-1}\hat{p})        \\   
\end{eqnarray*}
and
\begin{eqnarray*}
   bd^{\varphi_x} & = & \gamma(p)
                         (1+c^{-1}\bar{z}\cdot\hat{p})^{-1} \bar{z} \\
   a^{\varphi_x}  & = & \gamma (1+c^{-1}\bar{z}\cdot\hat{p})^{-2}
                        \big\{\bar{z}+c^{-1}\hat{p}
                        -c^{-2}\hat{p}\wedge(\bar{z}\wedge\hat{p})\big\}        \\ 
   b^{\varphi_x}  & = & \bar{z}b^{\varphi_t} \\
   c^{\varphi_x}  & = & \bar{z}\oplus c^{\varphi_t}\in\R^{3\times 3}.         
\end{eqnarray*}

Next we expand these kernels in powers of $c^{-1}$ (in this subsection we will
explain the argument leading to a formula for $\nabla_x\varphi$ in  detail and only give the formulas
for  the other fields). According to (\ref{SchrankeSupport})
 we can assume that the $p$-support
of $f(t, x, \cdot)$ is uniformly bounded in $x\in\R^3$ and $t\in [0, T]$, say
$f(t, x, p)=0$ for $|p|\ge P$. Thus we may suppose that $|p|\leq P$ in each of
the $p$-integrals, and hence also $|\hat{p}|=\gamma(p)|p|\leq |p|\leq P$
uniformly in $c$. It follows that
\begin{equation*}
   \hat{p}=(1-1/2c^{-2}p^2)p+{\cal O}(c^{-4}).
\end{equation*}   
For instance, for the kernel $bd^{\varphi_x}$ of $\varphi_{x, BD}$ this yields 
\begin{eqnarray*}
   bd^{\varphi_x} & = &  \bar{z}\gamma(p)(1+\frac{\bar{z}\cdot\hat{p}}{c})^{-1} 
   \\ & = & \bar{z}\Big[1-1/2c^{-2}p^2+{\cal O}(c^{-4})\Big]\bigg[1-c^{-1}(\bar{z}\cdot p)
            +c^{-2}(\bar{z}\cdot p)^2+c^{-3}\Big(1/2(\bar{z}\cdot p) p^2-(\bar{z}\cdot p)^3\Big)+ {\cal O}(c^{-4})\bigg] 
   \\ & = & \bar{z}\Big(1-c^{-1}(\bar{z}\cdot p)+c^{-2}\big((\bar{z}\cdot
             p)^2-1/2p^2\big)+c^{-3}\big((\bar{z}\cdot
             p)p^2-(\bar{z}\cdot p)^3\big)\Big)+{\cal O}(c^{-4}).
\end{eqnarray*}
If we choose $R_0>0$ such that $f^\circ(x, p)=0$ for $|x|\geq R_0$, then
\begin{eqnarray*}
   -(ct)^{-1}\int_{|z|= ct}\int_{|p|\leq P} {\cal O}(c^{-4})\mathbf{1}_{B(0,
    R_0)}(x+z)\,dp\,ds(z) & = & \bigg(ct \int_{|\omega|=1}\mathbf{1}_{B(0,
    R_0)}(x+ct\omega)\,ds(\omega)\bigg){\cal O}(c^{-4})
    \\ & = & {\cal O}(c^{-4})
\end{eqnarray*} 
by \cite[Lemma 1]{schaeffer:86}, uniformly in $x\in\R^3$, $t\in [0, T]$ and $c\ge 1$.
Therefore we arrive at
\begin{eqnarray}\label{gphiDT-expa}
  \lefteqn{\varphi_{x,BD}(t, x)}
  \nonumber \\
   & = & -c^{-2}(ct)^{-1}\int_{|z|=ct}\int \bar{z}
   \Big(1-c^{-1}(\bar{z}\cdot p)+c^{-2}\big((\bar{z}\cdot
   p)^2-1/2p^2\big)+c^{-3}\big((\bar{z}\cdot
   p)p^2-(\bar{z}\cdot p)^3\big)\Big)\times
  \nonumber \\
   & & \times f^\circ(x+z, p)\,dp\,ds(z) +{\cal O}(c^{-6}).
\end{eqnarray}
Concerning $\varphi_{x,a}$, we note that $f(t, x, p)=0$ for $|x|\ge R_0+TP=:R_1$.
Since, by distinguishing the cases $|x-y|\ge 1$ and $|x-y|\le 1$,
\[ \int_{|z|\leq ct} |z|^{-2}\,{\bf 1}_{B(0, R_1)}(x+z)\,dz
   =\int_{|x-y|\leq ct} |x-y|^{-2}\,{\bf 1}_{B(0, R_1)}(y)\,dy={\cal O}(1) \]
uniformly in $x\in\R^3$, $t\in [0, T]$, and $c\ge 1$, similar computations
as before show that
\begin{eqnarray}\label{gphia-expa}
   \varphi_{x,a}(x, t) & = & -c^{-2}\int_{|z|\leq ct} \int
   \Big(\bar{z}+c^{-1}\big[p-2(\bar{z}\cdot p)\bar{z}\big]+
   c^{-2}\big[3(\bar{z}\cdot p)^2\bar{z}-3/2p^2\bar{z}-(\bar{z}\cdot p)
   p\big] 
   \nonumber \\ & & \hspace{8em}
    c^{-3}\big[4\bar{z}(\bar{z}\cdot p)p^2-4(\bar{z}\cdot
   p)^3\bar{z}+(\bar{z}\cdot p)^2 p-p^2 p \big]\Big)
   \nonumber \\
   & & \hspace{8em} f(\hat{t}(z), x+z, p)\,dp\,\frac{dz}{|z|^{2}}+{\cal O}(c^{-6}).
\end{eqnarray}
In the same manner, elementary calculations using also (\ref{SchrankeFelder}) can be
carried out to get
\begin{eqnarray}\label{gphib-expa}
   \varphi_{x,b}(t, x) & = & -c^{-3}\int_{|z|\leq ct}\int \bar{z} 
   S(\varphi)f(\hat{t}(z), x+z, p)\,dp\,\frac{dz}{|z|}+{\cal O}(c^{-4})
   \\ \label{gphic-expa}
   \varphi_{x,c}(t, x) & = & -c^{-2}\int_{|z|\leq ct}\int \bar{z}\oplus
   \bigg(\bar{z}+c^{-1}\Big[-2(\bar{z}\cdot p)\bar{z}+p\Big]\bigg)
   \nonumber\\ 
   & & \hspace{8em} (\nabla_x\varphi)f(\hat{t}(z), x+z, p)\,dp\,\frac{dz}{|z|}+{\cal O}(c^{-4}),
   \\ \label{dtphiBD-expa}
   \varphi_{t,BD}(x, t) & = & -c^{-1}(ct)^{-1}\int_{|z|=ct}\int 
   \Big(1-c^{-1}(\bar{z}\cdot p)+c^{-2}\big((\bar{z}\cdot
   p)^2-1/2p^2\big)\Big) 
   \nonumber \\
   & & \hspace{8em}f^\circ(x+z, p)\,dp\,ds(z)
   +{\cal O}(c^{-4}),
   \\ \label{dtphia-expa}
    \varphi_{t,a}(x, t) & = & c^{-2}\int_{|z|\leq ct} \int
   \Big((\bar{z}\cdot p)+c^{-1}\big(p^2-2(\bar{z}\cdot p)\big)\Big) 
   f(\hat{t}(z), x+z, p)\,dp\,\frac{dz}{|z|^{2}}
   \nonumber \\
   & & \hspace{8em} +{\cal O}(c^{-4}),
   \\ \label{dtphib-expa} 
   \varphi_{t,b}(t, x) & = & -c^{-2}\int_{|z|\leq ct}\int 
   S(\varphi)f(\hat{t}(z), x+z, p)\,dp\,\frac{dz}{|z|}+{\cal O}(c^{-2}),
   \\ \label{dtphic-expa}
    \varphi_{t,c}(t, x) & = & -c^{-1}\int_{|z|\leq ct}\int 
   \bar{z}\cdot\nabla_x\varphi f(\hat{t}(z), x+z,
   p)\,dp\,\frac{dz}{|z|}+{\cal O}(c^{-2}),
   \\ \label{phi-expa}
   \varphi_S(t, x) & = &  -\frac{1}{c^2}\int_{|z|\leq
   ct}\int f(\hat{t}(z), x+z,
   p)\,dp\,\frac{dz}{|z|}+{\cal O}(c^{-4}).
\end{eqnarray}
Next we consider the data terms treating first the  gradient term.
\begin{eqnarray}
   \varphi_{x, D} & = & \nabla_x\partial_t\bigg(\frac{1}{4\pi}\int_{|\omega|=1}
   \varphi^0(x+ct\omega)\,d\omega)\bigg)+\nabla_x\bigg(\frac{t}{4\pi}
   \int_{|\omega=1}\varphi^1(x+ct\omega)\,d\omega\bigg)
   \nonumber \\
   & = & \partial_t\bigg(\frac{1}{4\pi}\int_{|\omega|=1}
   \nabla_x\varphi^0(x+ct\omega)\,d\omega)\bigg)+\frac{t}{4\pi}
   \int_{|\omega=1}\nabla_x\varphi^1(x+ct\omega)\,d\omega
   \nonumber \\    
   & = & I+II
\end{eqnarray}
Since $f_2(0, x, p)=0$ by (\ref{IC-LVP}), we have $\int f_2(0, x,
p)\,dp=0$. Thus we get from (\ref{IC-matched}), (\ref{VP}) and
(\ref{phi4-def})
\begin{eqnarray}
   \nabla_x\varphi^0(x) & = &c^{-2}\nabla_x\varphi_2(0, x)+c^{-4}\nabla_x\varphi_4(0, x)
   \nonumber \\
   & = & -\frac{1}{c^2}\int\int|z|^{-2}\bar{z}(1-\frac{p^2}{2c^2})f_0(0, x+z,
   p)\,dp\,dz+
   \frac{1}{2c^4}\int\bar{z}\partial_t^2\mu_0(0, x+z)\,dz.
   \nonumber
\end{eqnarray}
Using the formulas (\ref{formeli})-(\ref{formeliv}) below, we calculate
\begin{eqnarray}
   \lefteqn{\frac{-1}{c^2}\int_{|\omega|=1}\int\int|z|^{-2}\bar{z}\big(1-\frac{p^2}{2c^2}\big)f_0(0,
   x+ct\omega+z, p)\,dp\,dz\,d\omega}\hspace{2em}
   \nonumber \\
   & = & \frac{-1}{c^2}\int\int(1-\frac{p^2}{2c^2})f_0(0, y,
   p)\,dp\int_{|\omega|=1}|y-x-ct\omega|^{-3}(y-x-ct\omega)\,d\omega\,dy
   \nonumber \\
   & = & \frac{-4\pi}{c^2}\int_{|z|\ge ct}\int|z|^{-2}\bar{z}(1-\frac{p^2}{2c^2})f_0(0,
   x+z, p)\,dp\,dz,
   \nonumber \\
   \lefteqn{\frac{1}{2c^4}\int_{|\omega|=1}\int\bar{z}\partial_t^2\mu_0(0,
   x+ct\omega+z)\,dz\,d\omega}\hspace{2em}
   \nonumber \\
   & = &  \frac{1}{2c^4}\int\partial_t^2\mu_0(0,
   y)\int_{|\omega|=1}|y-x-ct\omega|^{-1}(y-x-ct\omega)\,d\omega\,dy
   \nonumber \\
   & = & \frac{2\pi}{c^4}\int_{|z|\ge
   ct}(\bar{z}-\frac{1}{3}(ct)^2|z|^{-2}\bar{z})\partial_t^2\mu_0(0,
   x+z)\,dz+\frac{4\pi}{3c^5t}\int_{|z|\leq ct}z\partial_t^2\mu_0(0, x+z)\,dz.
   \nonumber
\end{eqnarray}
Therefore we get
\begin{eqnarray}\label{gphiDi}
   \lefteqn{I=\partial_t\bigg(\frac{t}{4\pi}\int_{|\omega|=1}\varphi^0(x+ct\omega)\,d\omega\bigg)}\hspace{2em}
   \\
   & = & \partial_t\bigg\{\frac{-t}{c^2}\int_{|z|\ge ct}\int|z|^{-2}\bar{z}(1-\frac{p^2}{2c^2})f_0(0,
   x+z, p)\,dp\,dz
   \nonumber \\
   & &
   +\frac{t}{2c^4}\intgr\Big(\bar{z}-\frac{(ct)^2}{3}|z|^{-2}\bar{z}\Big)\dt^2\mu_0(0,
   x+z)\,dz+\frac{1}{3c^5}\intkl z\dt^2\mu_0(0, x+z)\,dz\bigg\}
   \nonumber \\
   & = & \frac{-1}{c^2}\int_{|z|\ge
   ct}\int|z|^{-2}\bar{z}\big(1-\frac{p^2}{2c^2}\big)f_0(0, x+z,
   p)\,dp\,dz+\frac{1}{2c^4}\int_{|z|\ge ct}\bar{z}\partial_t^2\mu_0(0,
   x+z)\,dz
   \nonumber \\
   & & -\frac{t^2}{2c^2}\int_{|z|\ge ct}|z|^{-2}\bar{z}\partial_t\mu_0(0,
   x+z)\,dz
   +\frac{1}{c^3t}\int_{|z|=ct}\int\bar{z}\Big(1-\frac{p^2}{2c^2}\Big) f_0(0,
   x+z, p)\,dp\,ds(z),
   \nonumber
\end{eqnarray}
note that several terms have canceled here. A similar calculation yields
\begin{eqnarray}\label{gphiDii}
   \lefteqn{II =  -\frac{t}{c^2}\int_{|z|\ge ct}|z|^{-2}\int\bar{z}
   \Big(\partial_t f_0+\frac{\dt f_2}{c^2}-\frac{p^2\dt f_0}{2c^2}\Big)(0,
   x+z, p)\,dp\,dz}
   \nonumber \\
   & & +\frac{t}{2c^4}\intgr\bar{z}\dt^3\mu_0(0,
   x+z)\,dz-\frac{t^3}{6c^2}\intgr|z|^{-2}\bar{z}\dt^3\mu_0(0, x+z)\,dz
   \nonumber \\
   & & +\frac{1}{3c^5}\intkl z\dt^3\mu_0(0, x+z)\,dz.
\end{eqnarray}
We need to examine the last term in (\ref{gphiDii}) more closely. Employing
the Vlasov-equation and integration by parts twice, we have
\begin{eqnarray}
  \label{gphiDiii}
     \lefteqn{\frac{1}{3c^5}\intkl z\dt^3\mu_0(0, x+z)\,dz =
     -\frac{t}{3c^4}\intgl\bar{z}(\bar{z}\cdot p) \dt^2\mu_0(0, x+z)\,ds(z)}
     \\
     & & -\frac{1}{3c^5}\intgl\int(\bar{z}\cdot p) p\dt f_0(0, x+z,
     p)\,dp\,ds(z)-
     \frac{1}{3c^5}\intkl\dt\big(\mu_0\nabla_x\varphi_2)(0, x+z)\,dz. 
     \nonumber
\end{eqnarray}
Using the support properties and the bounds
(\ref{SchrankeGeschwVP})--(\ref{SchrankeLVP}) as well as \cite[Lemma
1]{schaeffer:86} for the integrals over the boundary yields
  \begin{equation}
    \label{dgphib}
    \nabla\varphi_D={\cal O}(c^{-2}).  
   \end{equation}
Therefore, (\ref{SchrankeSupport}), (\ref{SchrankeFelder}), together with
(\ref{gphiDT-expa})--(\ref{gphic-expa}) and (\ref{dgphib}) gives
\begin{equation}
  \label{gphi-est}
  \nabla\varphi={\cal O}(c^{-2})
\end{equation}
and the error estimate in (\ref{gphic-expa}) is improved to 
\begin{eqnarray}
  \label{gphic-expa-im}
  \varphi_{x,c}(t, x) & = & -c^{-2}\int_{|z|\leq ct}\int \bar{z}\oplus
   \bigg(\bar{z}+c^{-1}\Big[-2(\bar{z}\cdot p)\bar{z}+p\Big]\bigg)
   \nonumber\\ 
   & & \hspace{8em} (\nabla_x\varphi)f(\hat{t}(z), x+z,
   p)\,dp\,\frac{dz}{|z|}+{\cal O}(c^{-6}).
\end{eqnarray}
We next claim that also 
\begin{equation}
  \label{bdtphi}
  \dt\phi={\cal O}(c^{-2}),
  \nonumber
\end{equation}
which gives the improved error estimate in (\ref{gphib-expa})
\begin{eqnarray}
  \label{gphib-expa-im}
  \varphi_{x,b}(t, x) & = & -c^{-3}\int_{|z|\leq ct}\int \bar{z} 
   S(\varphi)f(\hat{t}(z), x+z, p)\,dp\,\frac{dz}{|z|}+{\cal O}(c^{-6}).
\end{eqnarray}
Combining (\ref{gphi})--(\ref{gphia-expa}), (\ref{gphib-expa-im}),
(\ref{gphic-expa-im}) and
(\ref{gphiDi})--(\ref{gphiDiii}), we obtain the announced formulas (\ref{gphiext-rep})--(\ref{gphibd-rep}).

In order to prove the claim and to establish representation formulas we need to examine $\varphi_D$ and $\dt\varphi_D$
more closely. Although we only want to give an approximation up to order
$c^{-3}$, we have to take into account the terms coming from initial data of order
$c^{-4}$, too, because the associated homogenous fields have contributions
of lower order at least in the case of $\dt\phi_D$. Therefore we have to give
the full description of $\varphi_D$ and $\dt\varphi_D$. Since the calculations
are  similar to that already carried out, we restrict
ourself to give the formulas only:
\begin{eqnarray}
  \label{phiD-exakt}
  \varphi_D(t, x)  & = & 
  -\frac{1}{c^2}\intgr|z|^{-1}\Big(\mu_0+t\dt\mu_0\Big)(0, x+z)\,dz
  +\frac{1}{c^3}\intgl\int(\bar{z}\cdot p)f_0(0, x+z, p)\,dp\,dz 
  \nonumber \\
  & & +\frac{1}{2c^4}\intgr\int\Big(|z|^{-1}p^2f_0-(\bar{z}\cdot p)\dt f_0\Big)(0,
  x+z, p)\,dp\,dz
  \nonumber \\
  & & +\frac{t^2}{2c^2}\intgr\int|z|^{-2}(\bar{z}\cdot p)\dt f_0(0,
  x+z, p)\,dp\,dz 
  \nonumber \\ 
  & & +\frac{1}{c^5}\intkl\int\Big(-\dt f_2+\frac{p^2}{2}\dt
  f_0-\frac{|z|}{3}(\bar{z}\cdot p)\dt^2f_0\Big)(0, x+z, p)\,dp\,dz
  \nonumber \\
  & & -\frac{t}{c^4}\intgr\int|z|^{-1}\Big(\dt f_2+\frac{p^2}{2}\dt f_0\Big)(0,
  x+z, p)\,dp\,dz
  \nonumber \\
  & & -\frac{t}{2c^4}\intgr\int(\bar{z}\cdot p)\dt^2 f_0(0, x+z, p)\,dp\,dz
  \nonumber \\ 
  & & +\frac{t^3}{6c^2}\intgr\int|z|^{-2}(\bar{z}\cdot p)\dt ^2 f_0(0, x+z, p)\,dp\,dz. 
\end{eqnarray}
Taking into account the by now well known support properties and bounds from
(\ref{SchrankeGeschwVP})--(\ref{SchrankeLVP}) as well as e.g.
   \begin{eqnarray*}
      \frac{t^2}{c^2}\intgr|z|^{-2} \mathbf{1}_{B(0, M_0)}(x+z)\,dz &\leq&
      \frac{1}{c^4}\intgr\mathbf{1}_{B(0, M_0)}(x+z)\,dz\leq Mc^{-4}, \\
      \frac{1}{c^5}\intkl|z|\mathbf{1}_{B(0, M_0)}(x+z)\,dz & \leq & 
      \frac{t}{c^4}\intkl\mathbf{1}_{B(0, M_0)}(x+z)\,dz\leq Mc^{-4}
   \end{eqnarray*}
(\ref{phiD-exakt}) gives
\begin{equation}
  \label{phiD-approx}
  \varphi_D(t, x)   = 
  -\frac{1}{c^2}\intgr|z|^{-1}\Big(\mu_0+t\dt\mu_0\Big)(0, x+z)\,dz
  +\frac{1}{c^3}\intgl\int(\bar{z}\cdot p)f_0(0, x+z, p)\,dp\,dz + {\cal O}(c^{-4}).
\end{equation}
Combining (\ref{phi}), (\ref{phi-expa}) and (\ref{phiD-approx}) we obtain
\begin{equation}
  \label{phi-approx}
  \varphi =
  \varphi_\mathtt{ext}+\varphi_\mathtt{int}+\varphi_\mathtt{bd}+{\cal O}(c^{-4})
\end{equation}
with
\begin{eqnarray*}
  \varphi_\mathtt{ext}(t, x ) & = & -\frac{1}{c^2}\intgr|z|^{-1}\Big(\mu_0+t\dt\mu_0\Big)(0, x+z)\,dz,\\
  \varphi_\mathtt{int}(t, x) & = &  -\frac{1}{c^2}\int_{|z|\leq
   ct}\int f(\hat{t}(z), x+z,
   p)\,dp\,\frac{dz}{|z|}, \\
   \varphi_\mathtt{bd}(t, x) & = & \frac{1}{c^3}\intgl\int(\bar{z}\cdot p)f_0(0, x+z, p)\,dp\,dz.
\end{eqnarray*}
Differentiating (\ref{phiD-exakt}) w.r.t $t$ and applying  similar estimates yields
\begin{eqnarray}
  \label{dtphiD-approx}
  \dt\varphi_D(t, x) & = & 
  \frac{1}{c^2t}\intgl\mu_0(0, x+z)\,ds(z)
  +\frac{1}{c^2}\intgr\int|z|^{-2}(\bar{z}\cdot p) f_0(0, x+z, p)\,dp\,dz
  \nonumber \\
  & & -\frac{1}{c^3t}\intgl\int(\bar{z}\cdot p) f_0(0, x+z, p)\,dp\,ds(z)
  -\frac{1}{2c^4t}\intgl\int p^2 f_0(0, x+z, p)\,dp\,ds(z)
  \nonumber \\
  & & +\frac{t}{c^2} \intgr\int|z|^{-2}(\bar{z}\cdot p)\dt f_0(0, x+z,
  p)\,dp\,dz
  +{\cal O}(c^{-4}).
\end{eqnarray}
Combining (\ref{dtphi}), (\ref{dtphiBD-expa})--(\ref{dtphic-expa}) and
(\ref{dtphiD-approx}) we have
\begin{equation}
  \label{dtphi-approx}
  \dt\varphi =\varphi_{t, \mathtt{ext}}+\varphi_{t, \mathtt{int}}+\varphi_{t,
  \mathtt{bd}}  +{\cal O}(c^{-4})
\end{equation}
with
\begin{eqnarray*}
  \varphi_{t, \mathtt{ext}}(t, x) & = &
  \frac{1}{c^2}\intgr\int|z|^{-2}(\bar{z}\cdot p) (f_0+t\dt f_0)(0, x+z, p)\,dp\,dz,
  \\
  \varphi_{t, \mathtt{int}}(t, x) & = &  \frac{1}{c^2}\int_{|z|\leq ct} \int
   \Big((\bar{z}\cdot p)+c^{-1}\big(p^2-2(\bar{z}\cdot p)\big)\Big) 
   f(\hat{t}(z), x+z, p)\,dp\,\frac{dz}{|z|^{2}}
   \\
   & & -\frac{1}{c}\int_{|z|\leq ct}\int 
   \bar{z}\cdot\nabla_x\varphi f(\hat{t}(z), x+z,
   p)\,dp\,\frac{dz}{|z|},
   \\
   \varphi_{t, \mathtt{bd}}(t, x) & = & -\frac{1}{c^4t}\intgl\int(\bar{z}\cdot p)^2f_0(0, x+z, p)\,dp\,ds(z).
\end{eqnarray*}
Note, that because of  (\ref{dgphib}) and
(\ref{dtphiBD-expa})--(\ref{dtphic-expa}) and (\ref{dtphiD-approx}) we already
have 
\begin{equation}
  \label{dtphi-est}
  \dt\varphi={\cal O}(c^{-2}),  
\end{equation}
which in
turn gives
\begin{eqnarray*}
  \varphi_{t, b}(t, x) & = & {\cal O}(c^{-4})
  \\
  \varphi_{t, c}(t, x) & = &  -c^{-1}\int_{|z|\leq ct}\int 
   \bar{z}\cdot\nabla_x\varphi f(\hat{t}(z), x+z,
   p)\,dp\,\frac{dz}{|z|}+{\cal O}(c^{-4}),
\end{eqnarray*}
compare with  (\ref{dtphib-expa}) and
(\ref{dtphic-expa}).
\subsection{Some explicit integrals}
We point out some formulas that have been used in the previous sections.
For $z\in\R^3$ and $r>0$ an elementary calculation yields
\begin{equation}\label{formeli}
  \int_{|\omega|=1}|z-r\omega|^{-1}\,d\omega
  =\left\{\begin{array}{c@{\quad:\quad}c}
  4\pi r^{-1} & r\geq |z| \\ 4\pi |z|^{-1} & r\leq |z|
  \end{array}\right. .
\end{equation}
Differentiation w.r.t.~$z$ gives
\begin{equation}\label{formelii}
   \int_{|\omega|=1}|z-r\omega|^{-3}(z-r\omega)\,d\omega
   =\left\{\begin{array}{c@{\quad:\quad}c}
   0 & r>|z| \\ 4\pi |z|^{-2}\bar{z} & r<|z|
   \end{array}\right. .
\end{equation}
Similarly,
\[ \int_{|\omega|=1}|z-r\omega|\,d\omega
   =\left\{\begin{array}{c@{\quad:\quad}c}
   4\pi r+\frac{4\pi}{3}z^2 r^{-1} & r\geq |z| \\[1ex]
   4\pi |z|+\frac{4\pi}{3}r^2 |z|^{-1} & r\leq |z|
   \end{array}\right. , \]
and thus by differentiation
\begin{equation}\label{formeliv}
   \int_{|\omega|=1}|z-r\omega|^{-1}(z-r\omega)\,d\omega
   =\left\{\begin{array}{c@{\quad:\quad}c}
   \frac{8\pi}{3r}\,z & r>|z| \\[1ex] 4\pi\bar{z}
   -\frac{4\pi}{3}r^2|z|^{-2}\bar{z} & r<|z|\end{array}\right. .
\end{equation}
Finally, for $z\in\R^3\setminus\{0\}$ also
\begin{equation}\label{formelv}
   \int |z-v|^{-1}|v|^{-3}v\,dv=2\pi\bar{z}
\end{equation}
can be computed.


\bigskip\bigskip
\noindent
{\bf Acknowledgments:} The author is indebted to M.~Kunze 
and H.~Spohn for many discussions.


\begin{thebibliography}{99}

\bibitem{andcalrein}
 {\sc Andreasson H.\,\& Calogero S.\,\& Rein G.:}
 Global classical solutions to the spherically symmetric
 Nordstr{\"o}m-Vlasov-system,
 to appear in 
 {\em Math.~Proc.~Cam.~Phil.~Soc.~}
\bibitem{baukun}
 {\sc Bauer S.\,\& Kunze M.:} The Darwin approximation of the relativistic
 Vlasov-Maxwell system, to appear in
 {\em Ann.~H.~Poincar{\'e}}, ArXiv preprint {\tt math-ph/0401012}
\bibitem{calee}
 {\sc Calogero S.\,\& Lee H.:} The non-relativistic limit
 of the Nordstr\"{o}m-Vlasov system, 
 {\em Commun.~Math.~Sci.~}{\bf\,\,2}, 19-34 (2004)
\bibitem{carein}
 {\sc Calogero S.\,\& Rein G.:} On classical solutions of the
 Nordstr{\"o}m-Vlasov system, 
 {\em Comm.~Partial Differential Equations}{\bf\,\,28}, 1-29 (2003)
\bibitem{carein-II}
 {\sc Calogero S.\,\& Rein G.:} Global weak  solutions to  the
 Nordstr{\"o}m-Vlasov system, 
 {\em Differential Equations}{\bf\,\,204}, 323-338 (2004)
\bibitem{fried}
 {\sc Friedrich S.:} Global small solutions of the Vlasov-Nordstr{\"o}m
 system, ArXiv preprint {\tt math-ph/0404007}
\bibitem{glassey:96}
 {\sc Glassey R.T.:}{\em\,\,The Cauchy Problem in Kinetic Theory},
 SIAM, Philadelphia 1996
\bibitem{glstr}
 {\sc Glassey R.T.\,\& Strauss W.:} Singularity formation
 in a collisionless plasma could occur only at high velocities,
 {\em Arch.~Rational Mech.~Anal.~}{\bf 92}, 59-90 (1986)
\bibitem{horst:93}
 {\sc Horst E.:} On the asymptotic growth of the solutions
 of the Vlasov-Poisson system, {\em Math.~Methods Appl.~Sci.~}{\bf 16},
 75-85 (1993)
\bibitem{KR1}
 {\sc Kunze M.\,\& Rendall A.D.:} The Vlasov-Poisson system
 with radiation damping, {\em Ann.~H.~Poincar\'{e}}{\bf\,\,2},
 857-886 (2001)
\bibitem{KR2}
 {\sc Kunze M.\,\& Rendall A.D.:} Simplified models of electromagnetic and
 gravitational radiation damping, {\em Classical Quantum Gravity}{\bf\,\,18},
 3573-3587 (2001)
\bibitem{KS1}
 {\sc Kunze M.\,\& Spohn H.:} Slow motion of charges interacting
 through the Maxwell field, {\em Comm.~Math.~Phys.~}{\bf 212},
 437-467 (2000)
\bibitem{KS2}
 {\sc Kunze M.\,\& Spohn H.:} Post-Coulombian dynamics at order $c^{-3}$,
 {\em J.~Nonlinear Science}{\bf\,\,11}, 321-396 (2001)
\bibitem{lee}
 {\sc Lee H.:} The classical limit of the relativistic Vlasov-Maxwell system
 in two space dimensions, 
 {\em Math.~Methods Appl.~Sci.~}{\bf\,\,27}, 249-287 (2004)
\bibitem{lee-II}
 {\sc Lee H.:} Global existence of solutions of the Nordstr{\"o}m-Vlasov
 system in two space dimensions, ArXiv preprint {\tt math-ph/0312014}
\bibitem{lindner:91}
 {\sc Lindner A.:}{\em\,\,$C^k$-Regularit\"at der L\"osungen
 des Vlasov-Poisson-Systems partieller Differentialgleichungen},
 Diplom Thesis, LMU M\"unchen 1991
\bibitem{lions/perthame:91}
 {\sc Lions P.-L.\,\& Perthame B.:} Propagation of moments and regularity
 for the 3-dimensional Vlasov-Poisson system, {\em Invent.~Math.~}{\bf 105},
 415-430 (1991)
\bibitem{McO}
  {\sc McOwen R.C.:} The behavior of the Laplacian on weighted Sobolev
  spaces, {\em Comm.~Pure Appl.~Math.~}{\bf 32}, 783-795 (1979)
\bibitem{Nord}
 {\sc Nordstr{\"o}m G.:} Zur Theorie der Gravitation vom Standpunkt des
 Relativit{\"a}tsprinzip,
 {\em Ann.~Phys.~Lpz.~}{\bf 42}, 533 (1913)
\bibitem{pfaffelmoser:92}
 {\sc Pfaffelmoser K.:} Global classical solutions of the Vlasov-Poisson
 system in three dimensions for general initial data,
 {\em J.~Differential Equations}{\bf\,\,95}, 281-303 (1992)
\bibitem{rein}
 {\sc Rein G.:} Selfgravitating systems in Newtonian theory -- the
 Vlasov-Poisson system, in {\em Proc.~Minisemester on Math.~Aspects
 of Theories of Gravitation 1996}, Banach Center Publications {\bf 41},
 part I, 179-194 (1997)
\bibitem{ADR}
 {\sc Rendall A.D.:} The Newtonian limit for asymptotically flat solutions of
 the Vlasov-Einstein system, {\em Comm.~Math.~Phys.~}{\bf 163}, 89-112 (1994)
\bibitem{schaeffer:86}
 {\sc Schaeffer J.:} The classical limit of the relativistic Vlasov-Maxwell system,
 {\em Comm.~Math.~Phys.~}{\bf 104}, 403-421 (1986)
\bibitem{schaeffer:91}
 {\sc Schaeffer J.:} Global existence of smooth solutions to the
 Vlasov-Poisson system in three dimensions,
 {\em Comm.~Partial Differential Equations}{\bf\,\,16}, 1313-1335 (1991)
\bibitem{Sp}
 {\sc Spohn H.:}{\em\,\,Dynamics of Charged Particles and their Radiation Field},
 Cambrigde University Press, Cambridge 2004

\end{thebibliography}
\end{document}